\documentclass[pre,superscriptaddress,amsmath,amssymb,aps,twocolumn,floatfix]{revtex4-2}

\usepackage{graphicx}% Include figure files
\usepackage{dcolumn}% Align table columns on decimal point
\usepackage{bm}% bold math
\graphicspath{{figs/}}
\DeclareMathOperator\erfc{erfc}
\DeclareMathOperator\erf{erf}

\begin{document}

\preprint{AA}

\title{Dynamics and escape of active particles in a harmonic trap}

\author{Dan Wexler}
 \affiliation{Department of Physics Ben-Gurion University of the Negev, Beer-Sheva 84105, Israel}%Lines break automatically or can be forced with \\
\author{Nir Gov}%
\affiliation{Department of Chemical Physics,
The Weizmann Institute of Science, Rehovot 76100, Israel}%
\author{Kim \O. Rasmussen}
\affiliation{Theoretical division, Los Alamos National Laboratory, Los Alamos, NM 87454, USA}%
\author{Golan Bel$^1,$}
\affiliation{Department of Solar Energy and Environmental Physics, BIDR, Ben-Gurion University of the Negev}%
\affiliation{Center for Nonlinear Studies, Los Alamos National Laboratory, Los Alamos, NM 87545, USA}%
%\email{bel@bgu.ac.il}

\date{\today}% It is always \today, today,

\begin{abstract}
The dynamics of active particles is of interest at many levels and is the focus of theoretical and experimental research. There have been many attempts to describe the dynamics of particles affected by random active forces in terms of an effective temperature. This kind of description is tempting due to the similarities (or lack thereof) with systems in or near thermal equilibrium. However, the generality and validity of the effective temperature is not yet fully understood. Here, we studied the dynamics of trapped particles subjected to both thermal and active forces. The particles were not overdamped. Expressions for the effective temperature due to the potential and kinetic energies were derived, and they differ from each other. A third possible effective temperature can be derived from the escape time of the particle from the trap, using a Kramers-like expression for the mean escape time. We found that over a large fraction of the parameter space, the potential energy effective temperature is in agreement with the escape temperature, while the kinetic effective temperature only agrees with the former two in the overdamped limit. Moreover, we show that the specific implementation of the random active force, and not only its first two moments and the two point auto-correlation function, affects the escape time distribution.
\end{abstract}

\maketitle

\section{\label{sec:intro}Introduction}
Active systems, in which particles move due to non-thermal (active) forces, are of great current interest in statistical and biological physics \cite{ramaswamy2010mechanics,marchetti2013hydrodynamics,ramaswamy2017active}. Realizations of such systems include living systems \cite{Bernheim2018}, from cells to tissues ("active gels") \cite{prost2015active}, and synthetic systems \cite{dreyfus2005microscopic,palacci2013living}. In dense active systems, each particle is confined by its neighbors, leading to the formation of various condensed phases, such as motility-induced phase separation (MIPS) \cite{cates2015motility,zottl2016emergent}, active fluids and glasses \cite{berthier2013non}. The dynamics within these dense active systems \cite{Berthier2019,Janssen2019} is controlled by the rate at which particles undergo reorganization, and overcome local energy barriers. This dynamics has been explored in simulations of active glass \cite{henkes2011active,berthier2014nonequilibrium} and cell tissues \cite{bi2014energy,garcia2015physics,bi2016motility}. These events are driven by both thermal and active forces, and it is, therefore, appealing to describe them using a modified "effective temperature" ($T_{eff}$).

The notion of an effective temperature and its usefulness in describing the dynamics within non-equilibrium systems are still being explored and debated \cite{cugliandolo2011effective,wang2011,Puglisi2017}. Examples of systems for which this description was explored range from cells \cite{Betz2009,Park2010,Benisaac2011} to sheared granular matter \cite{Berthier2002,Ohren2004,nandi2019erratum}. Recently, several studies have indicated that the relaxation process in an active fluid and glass can be described as if driven by an effective temperature that has the form of the potential energy of an active particle in a confined harmonic potential \cite{nandi2017nonequilibrium,nandi2018random,cugliandolo2019effective} (which we denote as $T_x$). The underlying reason for this behavior is not fully understood.

These observations suggest that inside the dense systems, the activation of the particles, which allows them to escape their local confinement and rearrange themselves, is driven by an effective temperature $T_{eff}\simeq T_x$. Here, we wish to test this notion by studying a much simpler system of a single active particle confined in a finite harmonic potential (a recent study explored the active dynamics in a double-well potential \cite{caprini2019active} and in the overdamped limit of a general potential \cite{Woillez2019}). We simulate this process in a large range of parameters, including the underdamped regime, which is relevant for strong confinement. We find resonance effects due to the oscillations of the particle in the confining potential. By comparing the mean escape time to the predictions of Kramers’ escape theory, we find that the identification of $T_{eff}\simeq T_x$ gives a very good description of the active escape for a wide range of parameters. The differences between various implementations of the active force, all of which may have the same first moments and auto-correlation time, are investigated. Importantly, the limitations of the effective temperature approach and the conditions for its validity are explored in order to ensure that this simplifying approach to the non-equilibrium systems of active matter is not abused. These results, at the single particle level, can shed light on how activity drives the dynamics inside dense active systems.

Furthermore, the dynamics of trapped active particles, most simply modeled in a harmonic potential, was explored in recent years as a simplified model for the motion inside active glass \cite{fodor2016active}, gels \cite{ben2015modeling,razin2019signatures}, cells \cite{fodor2015activity,fodor2016nonequilibrium,ahmed2018active}, and tissues \cite{fodor2018spatial}. Trapped motile bacteria are another such system \cite{tailleur2009sedimentation,sevilla2019stationary}. Investigating the active escape from potential wells is, therefore, of general interest in a variety of different contexts.

The paper is organized as follows. In section \ref{sec:Theory}, we introduce the model and the implementations of the active force.
In section \ref{sec:anlytical}, we derive the expressions for the effective temperatures and the expressions for the mean escape time for an active force with long correlation times. In section \ref{sec:sim}, we present the results for the simulations of the escape time, including its dependence on the confining potential and active force characteristics. The results and their relation to the broader context of the dynamics of particles subjected to active forces are discussed in section \ref{sec:dis}, and the broader implications of the results for dynamics in non-equilibrium systems are summarized in section \ref{sec:sum}.

\section{\label{sec:Theory}Model and formulation of the problem}

The particle dynamics inside a trap is considered to be a one-dimensional dynamics within harmonic potential and under the action of both thermal and active noise.  
We also take into account the damping, which is assumed to be linearly proportional to the velocity. The thermal noise is implemented as a Gaussian white noise, and the active noise is implemented as a temporally correlated noise.
The equation of motion of the particle within the trap is \cite{ben2015modeling}:
\begin{equation}
m\ddot{x}=-\gamma\dot{x}-kx+f_{T}\left(t\right)+f_{a}\left(t\right).\label{eq:eomdim}
\end{equation}
$m$ is the mass of the particle, $\gamma$ is the friction coefficient, and $k$ is the harmonic constant. The thermal noise, $f_T(t)$, is characterized by:
\begin{align}\label{eq:thn}
    \langle f_T(t)\rangle &=0; \\
    \langle f_T(t)f_T(t')\rangle &=2k_BT\gamma\delta(t-t').\nonumber
\end{align}
The specific implementation of the active noise differs between different scenarios (one source or the collective action of many sources).
In all implementations, the auto-correlation function of the active noise is given by:
\begin{equation}\label{eq:anac}
    \left\langle f_{a}\left(t\right)f_{a}\left(t'\right)\right\rangle =\left\langle f_{a}^{2}\right\rangle \exp\left(-\frac{\left|t-t'\right|}{\tau_{on}}\right),
\end{equation}
and its first moment vanishes, $\langle f_a(t)\rangle=0$.
In some parts of this work, we also considered the dynamics in traps that are not harmonic. In these cases, the term $-kx$ in eq. \eqref{eq:eomdim} is replaced by the force corresponding to the actual potential (details are provided where these results are presented).

We note that we have several time scales in this problem; the natural
ones are the damping time, $t_{d}=m/\gamma$ and the oscillation
time, $1/\omega_{0}=\sqrt{m/k}$.
In addition, we have the active force correlation time, $\tau_{on}$ (the single source and $N$ source implementations of the active
force also have a mean period $\tau_{tot}\equiv\tau_{on}+\tau_{off}$).
If one considers a finite trap, namely the trapping potential is truncated at a certain point, $x_{esc}$, which if passed by the particle it escapes and its dynamics is not described by the above equations anymore, an additional relevant time scale is the thermal mean escape time, approximated by Kramers' formula,
\begin{equation}\label{eq:Kramers}
    \tau_{esc}^{thermal}=\tau_{0}e^{\Delta E(x_{esc})/k_{B}T},
\end{equation}
where $\Delta E(x_{esc})$
is the energy difference between the escape point and the bottom of the potential well, and $\tau_0$ is a coefficient with units of time \cite{Kramer1,Kramer2,Melnikov1991} (see Appendix \ref{App:tau0} for more details).

In order to simulate the dynamics of the particles, we used the non-dimensional version of eq. \ref{eq:eomdim} which reads,
\begin{equation}
\frac{d^2\tilde{x}}{d\tilde{t}^2}=-\frac{d\tilde{x}}{d\tilde{t}}-\alpha\tilde{x}+\phi\left(\tilde{t}\right)+\beta\varphi_s\left(\tilde{t}\right),\label{eq:eomnd}
\end{equation}
where $\tilde{x}=x/x_d$, the distance unit is $x_d\equiv\left(\sqrt{2mk_{B}T}/\gamma\right)$, and $\tilde{t}=t/t_d$.
 The confining potential stiffness is characterized by $\alpha=\omega_0^2t_d^2$,
and the dimensionless active force amplitude is characterized by $\beta=f_{0}/\left(mx_d/t_d^2\right)$ where $f_0$ is a constant (with dimensions of force) setting the amplitude of the active force.
$\phi\left(\tilde{t}\right)$ is a Gaussian white noise with $\langle\phi(\tilde{t})\rangle=0$ and $\langle\phi(\tilde{t})\phi(\tilde{t}')\rangle=\delta(\tilde{t}-\tilde{t}')$.
$\varphi_s\left(\tilde{t}\right)$ is the time-dependent part of the non-dimensional active force
($\varphi_s\left(\tilde{t}\right)=f_{a}\left(t_d\tilde{t}\right)/f_{0}$) and its characteristics depend on the active force implementation.
The index $s$ takes either an integer value representing the number of sources or the symbol $G$, which represents the Gaussian colored noise implementation of the active force.

The single source implementation of the active force consists of a sequence of "on" and "off" times. The duration of each "on"("off") period is drawn from an exponential distribution with the mean equal to $\tau_{on}$($\tau_{off}$). The fraction of "on' times is given by $p_{on}=\tau_{on}/\left(\tau_{on}+\tau_{off}\right)$. During "on" times, the amplitude of the active force is constant and equal to $f_0$ and its direction is randomly set for each period with equal probability for both sides (the active force is not biased). For implementations with any integer number of sources, the sources are assumed to be independent such that their "on"("off") states are not correlated, and their directions are also not correlated. The amplitude of each source during "on" times is $f_0/\sqrt{N}$ in order to ensure that the second moment of the force is the equal to that of the single source implementation.
The Gaussian implementation of the active force is done by considering it as an Ornstein-Uhlenbeck process with a correlation time given by $\tau_{on}$ and a second moment $\langle f_a^2\rangle$.
For all implementations of the active force, its first moment vanishes and the two-point correlation function is given by eq. \ref{eq:anac} with $\langle f_a(t_d\tilde{t})f_a(t_d\tilde{t}')\rangle=f_0^2p_{on}exp\left(-|\tilde{t}-\tilde{t}'|/(\tau_{on}/t_d)\right)$. The results will be presented using the force amplitude $\beta$ and/or the active "temperature" $T_a\equiv2\beta^2p_{on}T=\langle f_a^2\rangle/\left(k_{B}\gamma^2/m\right)$.

\begin{figure}[b]
\includegraphics[width=\linewidth,trim=5cm 0cm 6cm 0cm,clip]{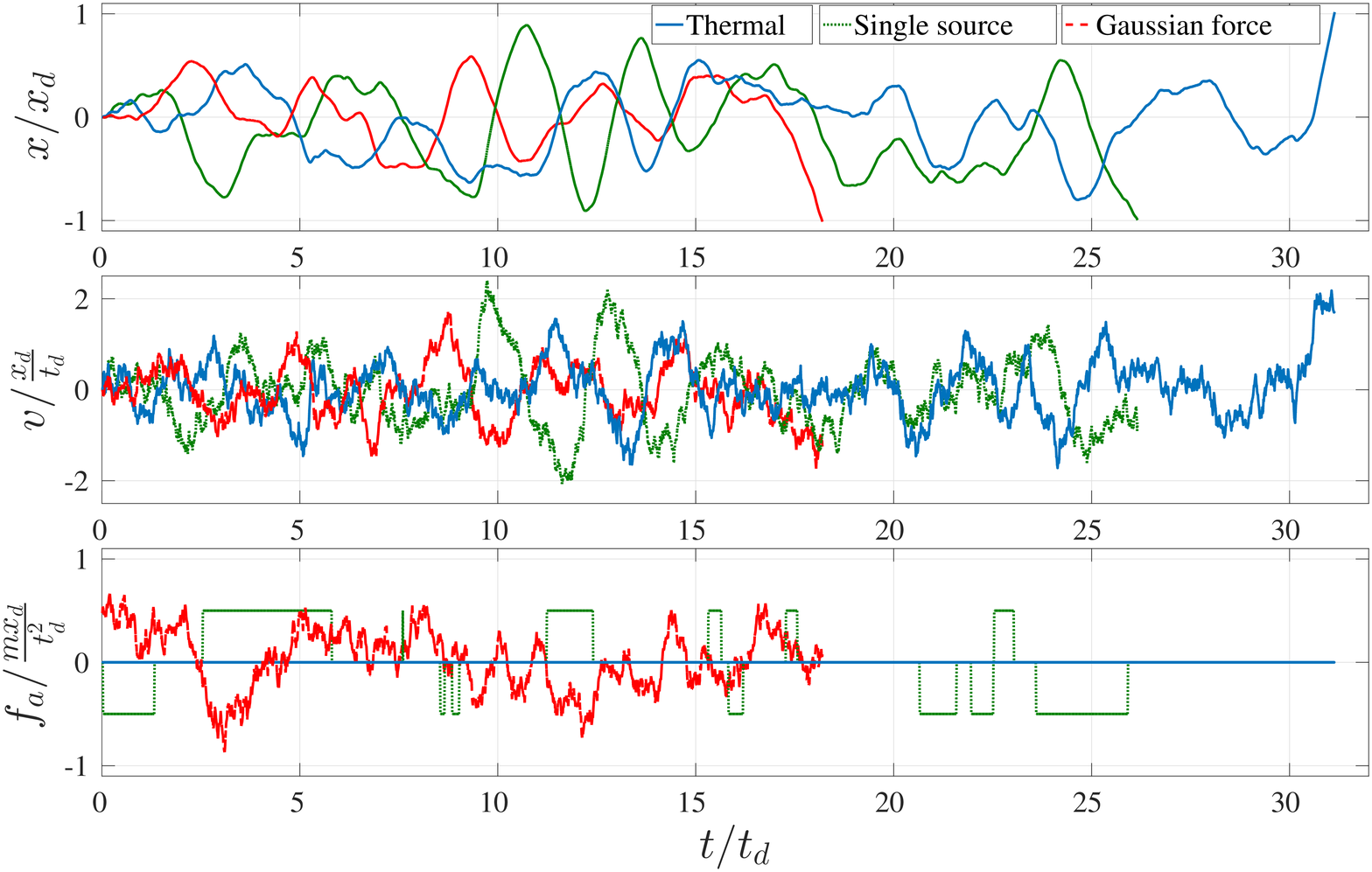}
\caption{\label{fig:etraj} Typical trajectories of the particle within a finite harmonic trap. The top panel shows the coordinate vs. time for a particle influenced by thermal noise alone (solid blue), a particle under the influence of both thermal noise and the Gaussian realization of the active force (dashed red), and a particle influenced by thermal noise and a single source realization of the active force (dotted green). The middle panel shows the velocity of the particle for the same scenarios, and the bottom panel shows the specific realization of the single source force and the Gaussian force. The trajectories were generated using simulations of the dynamics with the following parameters: $\alpha=3$, $\beta=0.5$, $x_{esc}=x_d$, and $p_{on}=0.5$. For these parameters, the restoring force due to the potential, at the escape point, is simply $\alpha m x_d/t_d^2=3 m x_d/t_d^2$, which is larger than the active force values shown in the figure (and realized in this trajectory).}
\end{figure}

In order to illustrate the dynamics considered here, we present typical trajectories in Fig. \ref{fig:etraj}.
\begin{figure}[b]
\includegraphics[width=\linewidth,trim=5cm 0cm 6cm 0cm,clip]{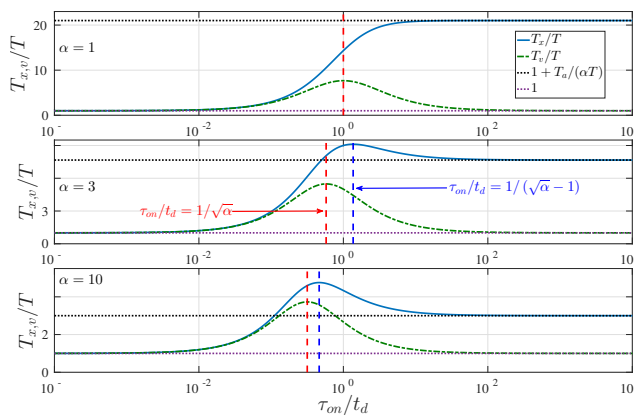}
\caption{\label{fig:T_eff_an}
The potential and kinetic energy effective temperatures for different values of $\tau_{on}$ (eqs. \eqref{eq:T_eff_x} and \eqref{eq:T_eff_v}). The different panels correspond to different stiffness values of the harmonic potential, $\alpha$. In all panels, the lower dashed horizontal line corresponds to $T_{eff}=T$ and the upper dashed line corresponds to the $lim_{\tau_{on}\to\infty}(T_x/T)=1+T_a/(\alpha T)$. The dashed vertical red line denotes the value of $\tau_{on}/t_d=1/\sqrt{\alpha}$ for which $T_v$ is maximal (the maximal value is $max(T_v)=1 + (T_a /T)\left(1/(2\sqrt{\alpha}+1)\right)$, and the dashed vertical blue line denotes the value of $\tau_{on}/t_d=1/\left(\sqrt{\alpha}-1\right)$ for which $T_x$ is maximal (the maximal value is $max(T_x)=1 + (T_a /T)\left(1/(2\sqrt{\alpha}-1)\right)$). In all panels, we used $T_a/T=20$ and $t_d=1$.
}
\end{figure}
\begin{figure}[b]
\includegraphics[width=\linewidth,trim=4cm 2cm 6cm 2cm,clip]{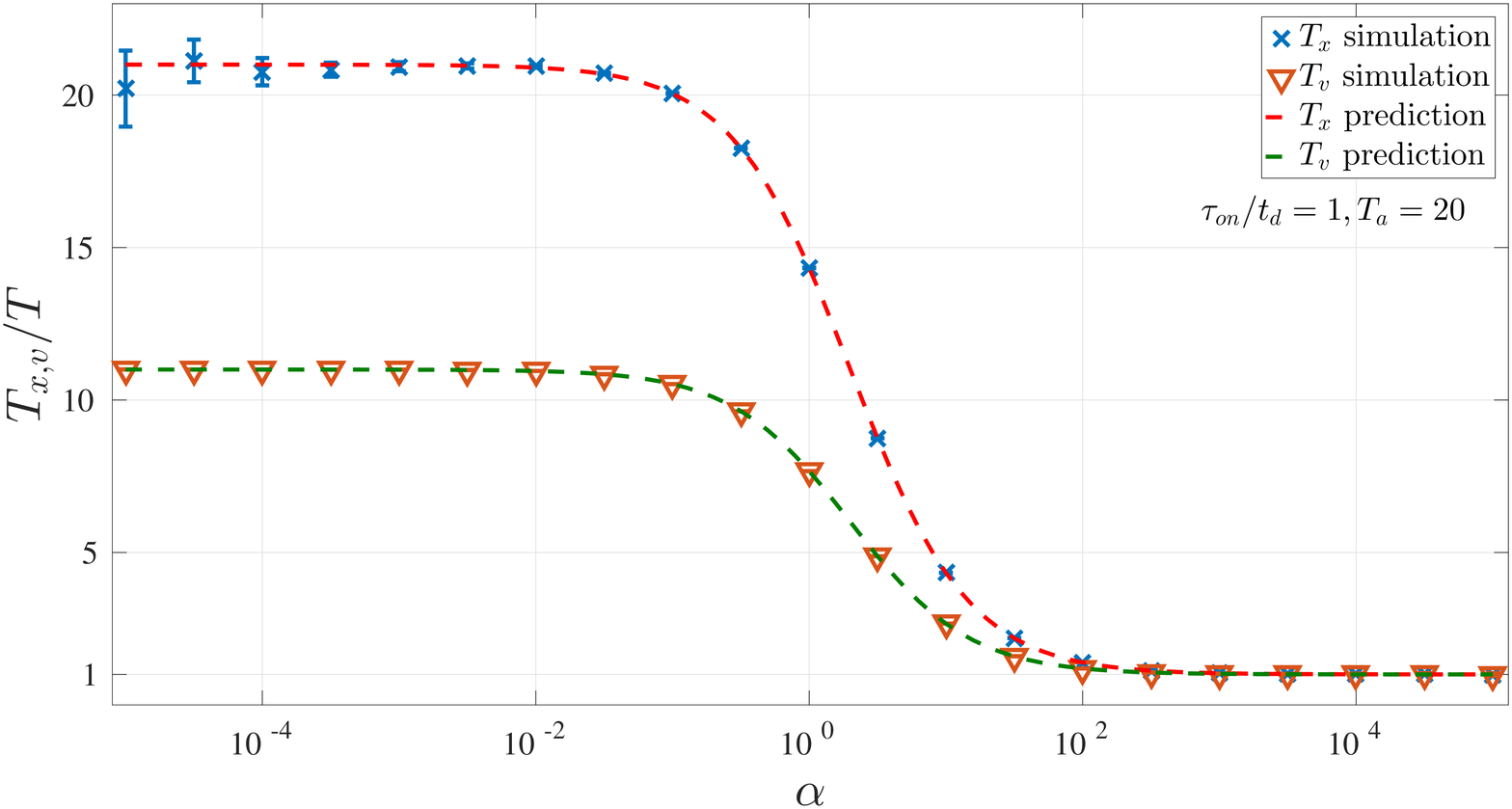}
\caption{\label{fig:simTeff}
Comparison of simulated and calculated effective temperatures. The potential and kinetic effective temperatures were derived from simulations of the dynamic of trapped particles using a wide range the harmonic potential stiffness parameter, $\alpha$. We used the Gaussian implementation of the active force with $T_a/T=20$ and $\tau_{on}=1$ for all the values of $\alpha$. There is an excellent agreement between the calculated and simulated second moments of the position and the velocity. For the small values of $\alpha$ the uncertainty (or spread of simulation results) is more apparent due to the broad range of $x$ values explored by the trapped particle.
}
\end{figure}
%
%\vskip 0.1\linewidth
\section{\label{sec:anlytical}  Analytical results}

The correlation time of the active force affects the nature of its effect on the dynamics of the particles. For short correlation times, it is tempting to characterize the effect using an effective temperature because the force statistics is explored before the particles escape the trap. The expressions for the effective temperatures are derived and discussed in subsection \ref{subsec:effT}. For long correlation times, particles may escape before the full statistics of the active force is experienced. Therefore, for long correlation times of the active force, we consider its effect on the particles by considering the modified confining potential. The characteristics of the escape time from the modified confining potential are derived and discussed in subsection \ref{subsec:effB}.

\subsection{\label{subsec:effT}Effective temperatures}

The system considered here is obviously out of equilibrium due to the active force, which is independent of the friction (does not obey any fluctuation-dissipation relation). Nevertheless, it is common to try and adapt the concept of temperature for the description of active matter properties. For the dynamics considered here, one may define effective temperatures based on the average potential energy (or the position fluctuations) or the average kinetic energy (or the velocity fluctuations). In addition, when the escape time of the particle is considered, one may try and identify the effective temperature in Kramers' sense, i.e., the effective temperature replacing the thermal temperature in Kramers' expression for the mean escape time.
We define the following effective temperatures:
\begin{align}
    T_x=& (k/k_B) \rm{lim}_{t\to\infty} \langle x^2\rangle;\\ \nonumber
    T_v=& (m/k_B) \rm{lim}_{t\to\infty} \langle v^2\rangle.
\end{align}
In order to derive the second moments of the velocity and the position, we write the formal solution for the coordinate (for simplicity, we set $x\left(t=0\right)=v\left(t=0\right)=0$)
\begin{align}\label{eq:implx}
x\left(t\right)=\frac{2}{m\omega}\intop_{0}^{t}e^{-\frac{\left(t-t'\right)}{2t_d}}\sinh\left(\frac{\omega\left(t-t'\right)}{2}\right)f\left(t'\right)\mathrm{d}t',
\end{align}
where $\omega=\sqrt{1-4\alpha}/t_d$, and $f\left(t\right)=f_{T}\left(t\right)+f_{a}\left(t\right)$. Taking the square of this expression and averaging over the thermal and the active noise using their characteristics (see eqs. \eqref{eq:thn} and \eqref{eq:anac}), we obtain for the potential energy effective temperature
\begin{equation}
T_x=T+T_a\frac{1+\tau_{on}/t_d}{1+\alpha\tau_{on}/t_d+t_d/\tau_{on}},\label{eq:T_eff_x}
\end{equation}
where the active "temperature" is defined as: $T_a=\left\langle f_{a}^{2}\right\rangle t_d^2/\left(m k_B\right)$.
A similar calculation for the second moment of the velocity yields for the kinetic energy effective temperature
\begin{equation}
T_v=T+T_a\frac{1}{1+\alpha\tau_{on}/t_d+t_d/\tau_{on}}.\label{eq:T_eff_v}
\end{equation}
Equations \eqref{eq:T_eff_x} and \eqref{eq:T_eff_v} show that $T_x\geq T_v$, or more precisely, $T_x-T=(T_v-T)\left(1+\tau_{on}/t_d\right)$. In the limit of $\tau_{on}\to0$, both effective temperatures converge to the thermal temperature, i.e., as expected, for a very short correlation time the active force does not contribute to the fluctuations of the trapped particle (note that the second moment of the active force is finite in this limit, which in combination with a vanishing correlation time, results in no contribution to the fluctuations of the trapped particles). In the opposite limit of a very large correlation time of the active force, $\tau_{on}\to\infty$, $T_v$ converges to the thermal temperature due to the fact that the "constant" active force just shifts the stationary point but does not affect the velocity fluctuations around the stationary point. $T_x$, on the other hand, is affected by the shift of the stationary point (since the fluctuations are around the stationary point rather than around the minimum of the potential, $x=0$ in our settings) and approaches the value of $lim_{\tau_{on}\to\infty}T_x=T+T_a/\alpha$.

The kinetic energy effective temperature has a maximum for a finite value of the correlation time, $\tau_{on}=t_d/\sqrt{\alpha}$. The potential energy effective temperature has a maximum only for $\sqrt{\alpha}>1$, and it is obtained for $\tau_{on}=t_d/(\sqrt{\alpha}-1)$. Note that a maximum of the second moment as function of the correlation time was also found for different systems \cite{Bel2013}.

The effective Kramers' temperature is unknown, and it is unclear if $T_x$ or $T_v$ are related to the escape properties in a similar way to the relation of the thermal temperature to the escape time in the absence of active force. The fact that these expressions were derived for the infinite time limit suggests that they may not be relevant for the description of an escape that occurs over shorter time scales than the convergence time to the asymptotic values of the respective second moments. The second moments for finite time include exponentially decaying terms with decay rates of $1/\tau_{on}$ and $\left(1\pm\sqrt{1-4\alpha}\right)/t_d$. Therefore, the values in \eqref{eq:T_eff_x} and \eqref{eq:T_eff_v}
are expected to be relevant for the escape process only if $\tau_{esc}\gg\max\left(\tau_{on},t_d/\left(\left(1-\sqrt{1-4\alpha}\right)\right)\right)$,
for $1>4\alpha$, or $\tau_{esc}\gg\max\left(\tau_{on},t_d\right)$.

The escape is defined by the coordinate and not by the velocity. Therefore, the fact that the process considered here is not overdamped suggests that the potential energy effective temperature, which describes the position fluctuations, would be more relevant (than the kinetic energy effective temperature) to the description of the escape process.

In Fig. \ref{fig:T_eff_an}, we present the effective temperatures (eqs. \eqref{eq:T_eff_x} and \eqref{eq:T_eff_v}) vs. the active force correlation time. The different panels correspond to different values of the harmonic potential stiffness, quantified by $\alpha=\left(\omega_0t_d\right)^2$. The limits of the effective temperatures for $\tau_{on}\to\infty$ are denoted by horizontal lines, and the values of $\tau_{on}$ that yield the maximal effective temperatures are denoted by vertical lines ($T_x$ has a maximum only for $\sqrt{\alpha}>1$, and therefore, its maximum is denoted only in the two lower panels).
We tested the validity of the analytically derived effective temperatures (eqs. \eqref{eq:T_eff_x} and \eqref{eq:T_eff_v}) for a wide range of potential stiffness values and found excellent agreement with the simulation results in all cases. The second moments, which were used to calculate the simulated effective temperatures, were derived from the simulated dynamics of trapped particles under the influence of thermal and active noise. The results are presented in Fig. \ref{fig:simTeff}.

\subsection{\label{subsec:effB}Modified potential}

Another way to characterize the effect of the active force is to consider the fact that the force modifies the potential and effectively reduces the barrier height.
Obviously, this effect is more significant when the correlation time of the force is long, and we expect this effect to be negligible for very short correlation times.
Considering the active force as constant results in a modification of the confining potential and asymmetric barrier height. The barrier height in the direction of the active force, ${\Delta \tilde{E}}^+=\Delta E-f x_{esc}$, is reduced while the barrier height in the opposite direction, ${\Delta \tilde{E}}^{-}=\Delta E+f x_{esc}$, is increased.

To find the modified (due to the modified confining potential) thermal escape time, we consider the asymmetric energy barrier and assume that the escape is the combination of two (escape to either side) rate processes, so
\begin{align}
{\tau_{esc}(f)}^{-1} & = \left(\tau_{esc}^{-}(f)\right)^{-1} +  \left(\tau_{esc}^{+}(f)\right)^{-1}\\
& = \tau_{1}^{-1} \left( e^{-\left(\frac{\Delta E + f x_{esc}}{k_B T}\right)} + e^{-\left(\frac{\Delta E - f x_{esc}}{k_B T}\right)} \right), \nonumber
\end{align}
where $\tau_1$ is a constant time from Kramers' expression and $\tau_{esc}^{\pm}$ are the mean escape times in the positive/negative directions, respectively. For consistency with Kramers' expression we must have $\tau_{esc}(f=0) = \tau_{esc}^{thermal}$, meaning $\tau_1 = 2 \tau_0$. It is important to note that for Kramers' expression to be valid, the force amplitude must be small enough such that $|\Delta \tilde{E}^{\pm}|>k_B T$. For large values of constant active force, we expect the escape to be almost ballistic (almost because the thermal noise is still acting on the particle).
We denote the ballistic time to escape as $t^*$ (neglecting the thermal noise, its value may be obtained by solving numerically the implicit equation, $$\frac{e^{-\frac{1}{2}(t^*/t_d)}\left(\sinh\left(\frac{t^*\omega}{2}\right)+\omega t_d\cosh\left(\frac{t^*\omega}{2}\right)\right)-\omega t_d}{m\omega\left(\omega^{2}t_d^2-1\right)/\left(4ft_d\right)}=x_{esc};$$ see eq. \eqref{eq:implx}), and it is expected, in the limit considered here of long correlation times of the active force, to be smaller than the thermal escape time. The resulting mean escape time for a given value of the active force is given by,
\begin{align}
    \tau_{esc}(f) = \begin{cases}
    \frac{\tau_{esc}^{thermal}}{\cosh\left(\frac{f x_{esc}}{k_B T}\right)} & |f|<<f_{cr},\\
    t^*(f) & |f|>>f_{cr},
    \end{cases}
\end{align}
where we defined $f_{cr}=\left(\Delta E-k_B T\right)/|x_{esc}|$

For the single source active force, the expected escape time depends on the ratio $\tau_{off}/\tau_{esc}^{thermal}$ and is given by:
\begin{widetext}
\begin{align}
\tau_{esc}\approx\begin{cases}
\tau_{esc}(f_0)+\left(1-p_{on}\right)\tau_{off} & \tau_{off}\ll\tau_{esc}^{thermal},\\
\tau_{esc}^{thermal}\cdot\left(p_{on}\tau_{esc}(f_0)/\tau_{esc}^{thermal}+\left(1-p_{on}\right)\right) & \tau_{off}\gg\tau_{esc}^{thermal}.
\end{cases}
\end{align}
\end{widetext}
There is a probability $(1-p_{on})$ that the active force is zero at the beginning. Therefore, for $\tau_{off}\ll\tau_{esc}^{thermal}$, the expected escape time is the weighted sum of waiting time for the active force to be non-zero ($\tau_{off}$) and the escape time over the reduced barrier height. For $\tau_{off}\gg\tau_{esc}^{thermal}$, the expected escape time is the weighted sum of the thermal escape time and escape time out of the modified trap. In many cases, the ballistic escape time, $t^*$, may be neglected relative to the terms involving the thermal escape time.

For the continuous Gaussian active force, with a correlation time much longer than the escape time, we have to consider the Gaussian distribution of the initial force amplitude.
In this implementation of the active force, the large amplitude only implies a large variance of the active force, but weaker forces are still possible (though with a smaller probability due to the broadening of the distribution).
The mean escape time may be written as
\begin{widetext}
\begin{align}\label{eq:intfton}
    \tau_{esc} & \approx\intop_{-\infty}^{\infty}p\left(f\right)\tau_{esc}\left(f\right)\,\mathrm{d}f
    \approx\frac{\tau_{esc}^{thermal}}{\sqrt{2\pi\left\langle f_{a}^{2}\right\rangle }}\intop_{-f_{cr}}^{f_{cr}}\frac{e^{-\frac{f^{2}}{2\left\langle f_{a}^{2}\right\rangle}}}{\cosh{\left(\frac{f x_{esc}}{k_B T}\right)}\,}\mathrm{d}f.
\end{align}
\end{widetext}
In the last expression on the RHS, we neglected $t^*$ for large force values. The integral can be numerically calculated or approximated by expanding the denominator, whose argument is small, and taking the integration limits to infinity (the denominator and the Gaussian kernel ensure that the approximation is valid). The approximated mean escape time is given by
\begin{widetext}
\begin{equation}\label{eq:metlton}
    \tau_{esc}\approx \tau_{esc}^{thermal}\sqrt{\frac{\pi T}{2 T_a}}\frac{ x_d}{ x_{esc}}\exp{\left(\frac{T}{2T_a}\left(\frac{x_d}{x_{esc}}\right)^2\right)}\erfc{\left(\sqrt{\frac{T}{2T_a}}\frac{x_d}{x_{esc}}\right)},
\end{equation}
\end{widetext}
where $\erfc{(x)}\equiv 1-\erf{(x)}$ is the complementary error function. This approximation is expected to provide an upper limit for the mean escape time and results in an analytical expression.

\section{Simulation results}\label{sec:sim}

The escape times and other characteristics of the dynamics inside the trap
were simulated under various parameters. The simulations were done by numerically integrating (Euler or second order Runge-Kutta where needed in order to verify the convergence of the results \cite{Milshtein1994}) the non-dimensional equation of motion, eq. \eqref{eq:eomnd}.

All the simulations started with the particle at rest at the bottom of
the trap $\tilde{x}\left(0\right)=\dot{\tilde{x}}\left(0\right)=0$.
The time unit in the non-dimensional dynamics is simply $t_d$.
Equation \eqref{eq:eomnd} shows that except for the active force, all the other aspects of the dynamics inside the trap are controlled by the parameter $\alpha$ (which is proportional to $k$). The escape process is also dictated by the non-dimensional escape point, $\tilde{x}_{esc}=x_{esc}/x_d$.%\frac{\sqrt{2mk_{B}T}}{\gamma}$.

The effects of the active force on trapped particles may be seen in the probability density functions (PDFs) of the position and the velocity of the particles. We studied the PDFs for different characteristics of the confining potential and the active force.  The PDFs also depend on the implementation of the active force (we obtained the PDFs for the single source and the Gaussian implementations of the active force). We found that the PDFs may show significant deviations from a Gaussian distribution, especially for long correlation times of the active force. In all cases, we found an excellent agreement between the calculated second moments (eqs. \eqref{eq:T_eff_x} and \eqref{eq:T_eff_v}) and those obtained from the simulations. For particles that can escape the confining potential, i.e., when we set absorbing boundary conditions for $|x|=x_{esc}$, the PDFs change considerably as do the second moments. Therefore, it is not obvious that the effective temperatures are relevant for the description of the escape process. The PDFs and their detailed discussion are provided in Appendices \ref{app:A} and \ref{app:B}.

The mean escape time, which is the focus of this work, depends on the characteristics of the trap, the thermal noise, and the active force.
In particular, the active force may be implemented using either a single source, several sources, or the continuous version, which is a Gaussian colored noise. The correlation time and the second moment can be set equal in all the implementations.
In subsection \ref{ss:met_ebh}, we present the mean escape time for different values of the confining potential stiffness and different escape points (both of which modify the energy barrier). In subsection \ref{ss:met_ton}, we present the effects of the active force correlation time on the mean escape time for different amplitudes and implementations of the active force. In subsection \ref{ss:met_ns}, we study the effects of the number of sources generating the active force on the mean escape time, and the probability density function of the escape time is studied in subsection \ref{ss:et_pdf}.
In subsection \ref{ss:met_nh}, we study the mean escape time from non-harmonic confining potentials.

\subsection{Effects of the barrier height on the mean escape time}\label{ss:met_ebh}

\begin{figure}[b]
\includegraphics[width=\linewidth,trim=4cm 0cm 6cm 3.5cm,clip]{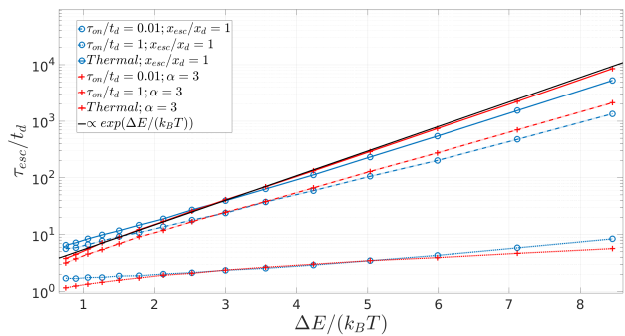}
\caption{\label{fig:metvsb}
Mean escape time vs. potential well barrier height for a Gaussian active force. The barrier height is controlled through either the potential's stiffness $\alpha$ (blue) or the escape point $x_{esc}$ (red) while the other is held constant. Line styles represent active force parameters where solid lines correspond to the thermal case (no active force) and the dashed and dotted lines correspond to active forces with $\tau_{on}/t_d = 0.01$ and $\tau_{on}/t_d = 1$, respectively. The solid black line is the Kramers’ predicted mean escape time, for reference. In all simulations with an active force, a value of $T_a/T = 20$ was used}
\end{figure}
First we examined the dependence of the mean escape time on the energy barrier. For the thermal case, Kramers' expression shows an exponential dependence on the magnitude of the energy barrier, regardless of what sets the barrier's height (the barrier’s height may vary by varying $x_{esc}$ or $\alpha$). The other characteristics of the potential well and the dynamics affect the pre-factor, $\tau_0$ (eq. \eqref{eq:Kramers}). In the case of an active force, there is no guarantee that a similar relation will hold even if one replaces the thermal temperature with an effective temperature. Moreover, the effective temperature is not uniquely defined because there is no fluctuation dissipation relation.

In Fig. \ref{fig:metvsb}, we present the mean escape time versus the barrier height. The barrier height was modified by varying either the escape point (red), $x_{esc}$, or the stiffness of the confining potential (blue), $\alpha$.
The different line styles refer to the thermal case and two different correlation times of the active force. The black line represents the exponential dependence to guide the eye. The longer correlation time of the active force reduces the mean escape time for all heights of the energy barrier. For a fixed $\alpha$, the dependence of the mean escape time on the energy barrier height is still exponential, but the slope depends on the active force characteristics. For the case of varying $\alpha$, a deviation from exponential dependence can be seen even for the thermal case. This deviation is due to the dependence of the pre-factor in Kramers' expression, eq. \eqref{eq:Kramers}, on $\alpha$. For more details, see Appendix \ref{App:tau0}.

\subsection{Effects of the active force correlation time on the mean escape time}\label{ss:met_ton}

In order to explore the high dimension parameter space, we decided to focus on several cross-sections of this space where we either set the ratio $p_{on}=\tau_{on}/\tau_{tot}$ constant or we kept $\tau_{tot}=\tau_{on}+\tau_{off}$ constant. For the first choice, varying $\tau_{on}$ implies varying $\tau_{off}$ by the same multiplicative factor. This, in turn, implies that larger "on" times are always accompanied by larger "off" times. For the explicit source implementations (i.e., single source or any small integer number of sources), this is reflected in larger "off" times, which affect the mean escape time if the thermal force is insufficient to trigger an escape during an "off" time. For the continuous force, only the ratio $p_{on}$ and the correlation time $\tau_{on}$ affect the active force characteristics.
\begin{figure*}[b]
\includegraphics[width=\linewidth]{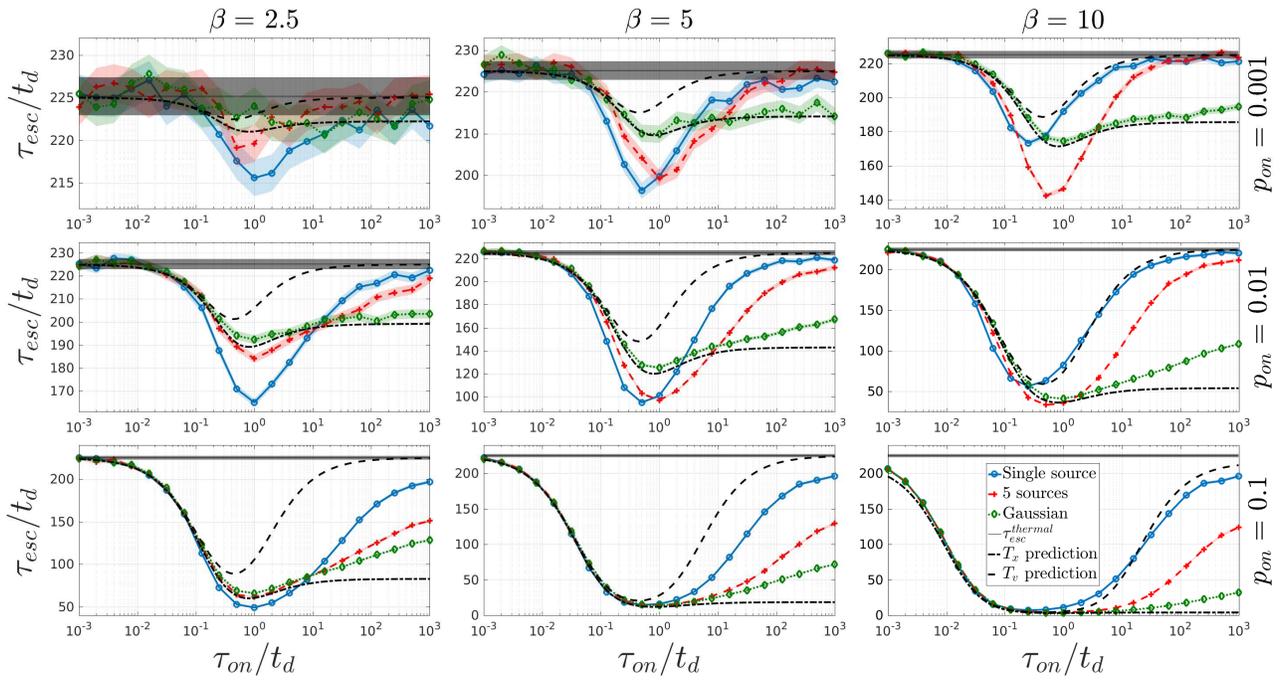}
\caption{\label{fig:tesc_DC_imp}
Mean escape time vs. the correlation time of the active force. The three rows correspond to the three indicated values of $p_{on}=\tau_{on}/\tau_{tot}$ (0.001, 0.01, and 0.1). The three columns correspond to the indicated force amplitude, $\beta$, (2.5, 5, and 10). In each panel, the curves show the mean escape time for the three different force implementations (single source, 5 sources, and the Gaussian) and the mean escape times expected from Kramers' expression with the effective temperatures, $T_x$ or $T_v$. The solid black line in each panel represents the thermal mean escape time. The shaded area represents the 95\% confidence interval. The confidence interval and the mean were derived from $4\times 10^4$ simulated trajectories. For all the simulations, $x_{esc}=x_d$ and $\alpha=5$.}
\end{figure*}
\begin{figure}[b]
\includegraphics[width=\linewidth,trim=4cm 0cm 6cm 3.5cm,clip]{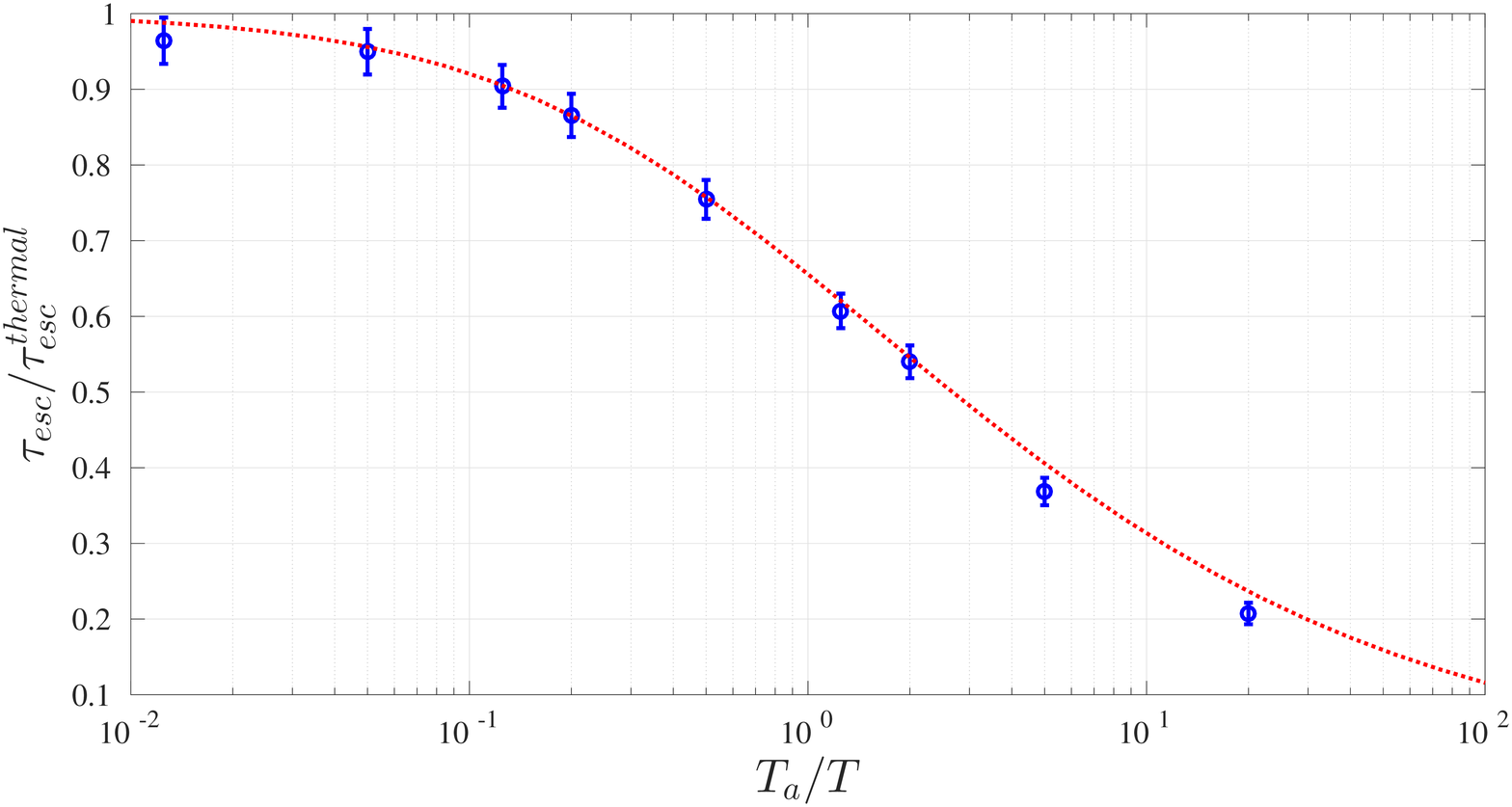}
\caption{\label{fig:met_lton}
Comparison of the simulated mean escape time with the theoretical approximation (eq. \eqref{eq:metlton}) for long correlation times of the active force. The simulations are based on the Gaussian implementation of the active force with $\tau_{on}=4 \times 10^4 t_d$. The confining potential parameters are:  $\alpha = 5, x_{esc} = x_d$.}
\end{figure}
We start by considering the constant $p_{on}$ case. In Fig. \ref{fig:tesc_DC_imp}, we present the mean escape times for different implementations of the active force: single source, five sources, and the Gaussian force. For each implementation, three different values of the ratio $p_{on}$ were considered, $p_{on}=10^{-3}$, $10^{-2}$, and $10^{-1}$. For each of the nine parameter sets, we considered three different amplitudes of the active force, $\beta=2.5$, $5$, and $10$.
The results show that for the single source and the five source implementations, there is a clear optimal correlation time that yields the shortest escape time. Considering the fact that increasing the correlation time results in increasing "off" times of the active force (because $p_{on}$ remains constant), this behavior is easily understood. We also notice that for short and long correlation times (relative to the optimal correlation time), the larger active force amplitude becomes less effective. For the Gaussian force, we still see the optimal correlation time despite the fact that there is no time during which the active force is "off". This behavior is explained by the match (resonance) between the correlation time of the active force and the time scale corresponding to the natural dynamics of the particle in the trap in the absence of an active force. Similar behavior was observed for a model describing the effect of stochastic wind stress on surface ocean currents \cite{Bel2013}.

\begin{figure*}[b]
\includegraphics[width=\linewidth]{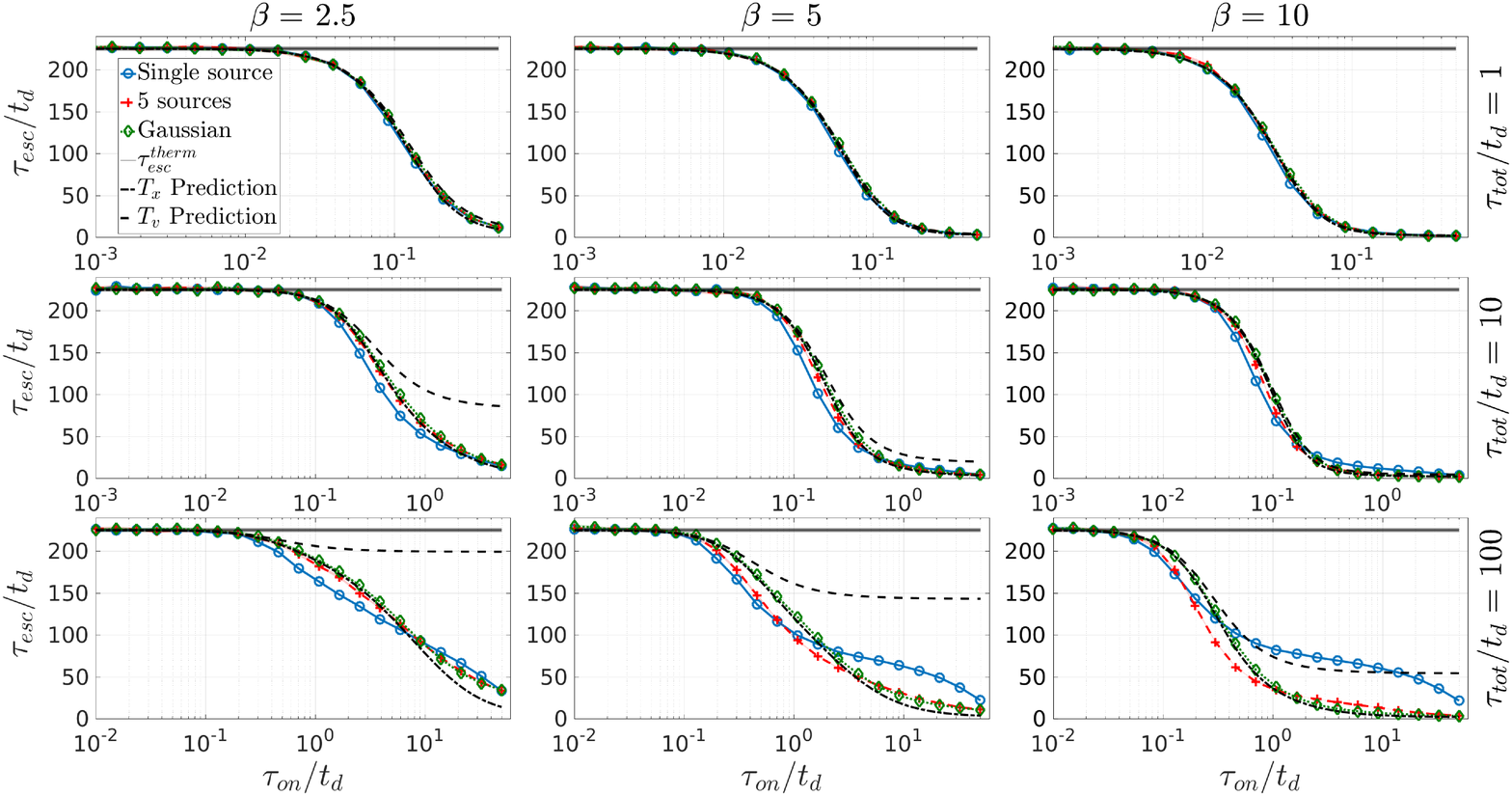}
\caption{\label{fig:tesc_cttot_imp}
Mean escape time vs. the correlation time of the active force. The three rows correspond to the three indicated values of $\tau_{tot}$ (1, 10, and 100). The three columns correspond to the three amplitudes of the active force, $\beta$ (2.5, 5, and 10). The different curves in each panel correspond to three implementations of the active force (single source, 5 sources, and the Gaussian) and to the expected mean escape times according to Kramers' expression with the effective temperatures $T_x$ or $T_v$. The solid black lines present the thermal escape time. The shaded areas (too narrow to see in most cases) present the 95\% confidence interval. The confidence interval and the mean were derived from $4\times 10^4$ simulated trajectories. For all the simulations, $x_{esc}=x_d$ and $\alpha=5$. }
\end{figure*}

For a short correlation time of the active force, the escape time approaches the thermal escape time due to the inefficiency of the active force, which can change sign before the particle escapes. It is important to note that the second moment of the active force approaches zero for short correlation times (unlike white noise whose second moment is proportional to a delta function and, therefore, has a finite contribution even in the limit of a vanishing correlation time). For long correlation times of the active force, there is a saturation but not to the thermal escape time. For the single source and the five source implementations of the active force, the convergence in the long correlation time limit is toward the escape time in the presence of a constant active force but also accounting for the probability that the force is "off" and there is a waiting time before the particle starts its active escape process. For the Gaussian force implementation, the convergence is toward the escape time in the presence of a constant active force but accounting for the weight of the possible active force amplitudes (see subsection \ref{subsec:effB}).

 The expected mean escape times according to Kramers' expression with the effective temperatures are shown in each panel. For short correlation times, the differences between the different implementations are small. The value of $\tau_{on}$ corresponding to the minimal mean escape time varies between the different implementations of the active force. For correlation times that are much smaller than the mean escape time, Kramers' expression with the potential energy effective temperature, $T_x$, shows good agreement with the results for the Gaussian implementation of the active force. For correlation times that are longer than the mean escape time, there is a very broad range between the $T_x$ and $T_v$ Kramers' expressions, and the simulation results are within this range. The minimal mean escape time for the single source and the five sources is not well captured by Kramers' expressions.
 
 In Fig. \ref{fig:tesc_cdc}, we present the same results of Fig. \ref{fig:tesc_DC_imp} but organized to allow an easy comparison between the different active force amplitudes.
 
 For long correlation times, we derived an approximation of the mean escape time, eq. \eqref{eq:metlton}. In Fig. \ref{fig:met_lton}, we compare the mean escape time derived from simulating the dynamics of particles escaping the trap with the analytical expression. The simulations are based on the Gaussian implementation of the active force with different amplitudes and correlation times given by $\tau_{on}=4 \times 10^4 t_d$. The confining potential parameters are: $\alpha = 5, x_{esc} = x_d$. The values of $T_a/T$ presented correspond to the values used in the different panels of Fig. \ref{fig:tesc_DC_imp}. Note that larger differences between the approximation of eq. \eqref{eq:metlton} and the simulation results for larger $T_a$ are expected due to the fact that the exponential term in eq. \eqref{eq:intfton} is wider and values of the active force for which the expansion of the denominator near zero is less accurate contribute to the integral. We also verified a convergence of the mean escape time for longer correlation times, the fact that the mean escape time is independent of the potential stiffness, $\alpha$, and the dependence of the mean escape time in the escape point, $x_{esc}$.

In Fig. \ref{fig:tesc_cttot_imp}, we present the mean escape time for the case of constant $\tau_{tot}$. For the single source and the five source implementations, keeping $\tau_{tot}$ constant implies that increasing the active force correlation time is accompanied by a decrease in the "off" times. For the Gaussian implementation, increasing $\tau_{on}$ implies increasing the active force amplitude, which is proportional to $p_{on}=\tau_{on}/\tau_{tot}$. Therefore, there is no optimal correlation time, and the mean escape time monotonically decreases with the increase in the correlation time for the active force implementations, for all values of $\tau_{tot}$ and active force amplitudes. Kramers' expressions for the mean escape time, based on the effective temperatures, are also presented. The expression based on $T_x$ shows good agreement with the Gaussian implementation of the active force in most cases. In Fig. \ref{fig:tesc_cttot}, we present the same results but organized to enable easier comparison between the different active force amplitudes.

In Fig. \ref{fig:tesc_comp}, we present the comparison between the simulated mean escape times and those calculated using Kramers' expressions and the effective temperatures, $T_x$ and $T_v$, for two sets of parameters and for the case of constant $\tau_{tot}$. The behavior described above can be clearly seen in this figure.

\subsection{Effects of the number of sources on the mean escape time}\label{ss:met_ns}

\begin{figure}[b]
\includegraphics[width=\linewidth,trim=4cm 1cm 4cm 2cm,clip]{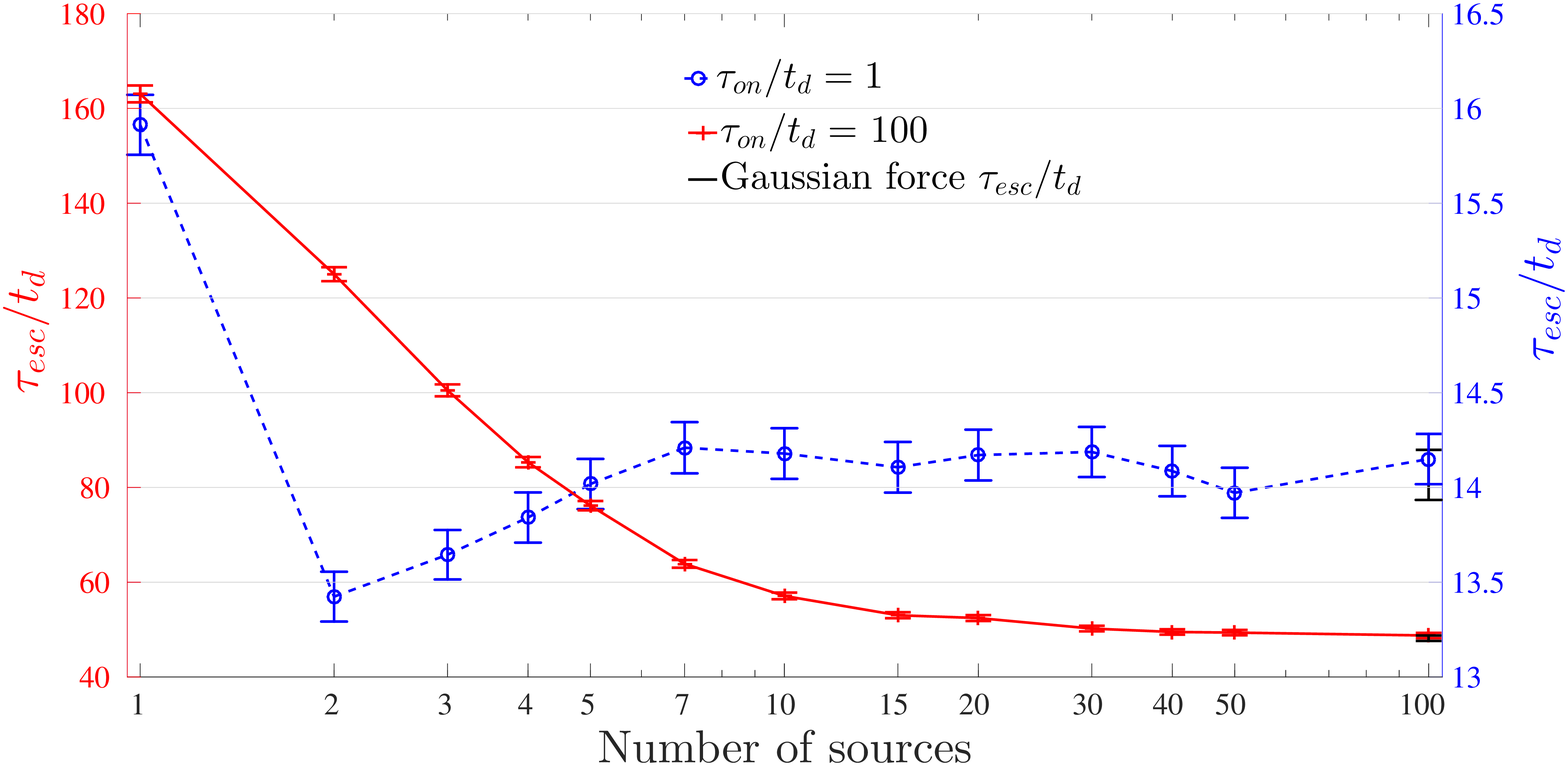}
\caption{\label{fig:tesc_ns}
Mean escape time vs. the number of sources used in the implementation of the active force. The different lines correspond to the denoted active force correlation times. The active force amplitude is, $\beta=2.5$, the confining potential stiffness is, $\alpha=5$, the mean fraction of activity time is, $p_{on}=0.1$ and the escape point is, $x_{esc}=x_d$. The error bars represent the 95\% confidence interval for the mean escape times derived from the simulation of $4\times10^4$ trajectories. The black symbols and error bars represent the mean escape time and the associated uncertainty for a Gaussian force.}
\end{figure}

As mentioned above, the Gaussian implementation of the active force is the limit of having many independent sources (unsynchronized). To better illustrate this, we present in Fig. \ref{fig:tesc_ns} the mean escape time versus the number of sources and the limiting values derived from the Gaussian implementation of the active force. The error bars denote the 95\% confidence interval. The fact that the mean escape time depends on the number of sources emphasizes the fact that the active force and its effects are not fully characterized by the correlation time and the second moment. These, in turn, imply that any description based on effective temperatures will be limited. For the long correlation time, $\tau_{on}=100t_d$, the mean escape time monotonically decreases with the number of sources. This decrease is due to the fact that the more sources there are, the less likely it is to have a time with no active force. For the shorter correlation time, $\tau_{on}=t_d$, one sees that the maximal mean escape time is found for the single source; there is a sharp decrease for two sources, and then almost a monotonic increase (within the uncertainty limits) toward the limit of the Gaussian force implementation. This behavior may be explained by the fact that when two sources act in the same direction, the net force is larger than the force of a single source (because the amplitude is only divided by $\sqrt{2}$). The probability of having all the sources acting in the same direction decreases with the number of sources; therefore, for more than two sources, there is an increase in the escape time until it saturates to the mean escape time corresponding to the Gaussian force implementation.

\subsection{The probability density function of the escape time}\label{ss:et_pdf}

\begin{figure*}[b]
\includegraphics[width=\linewidth]{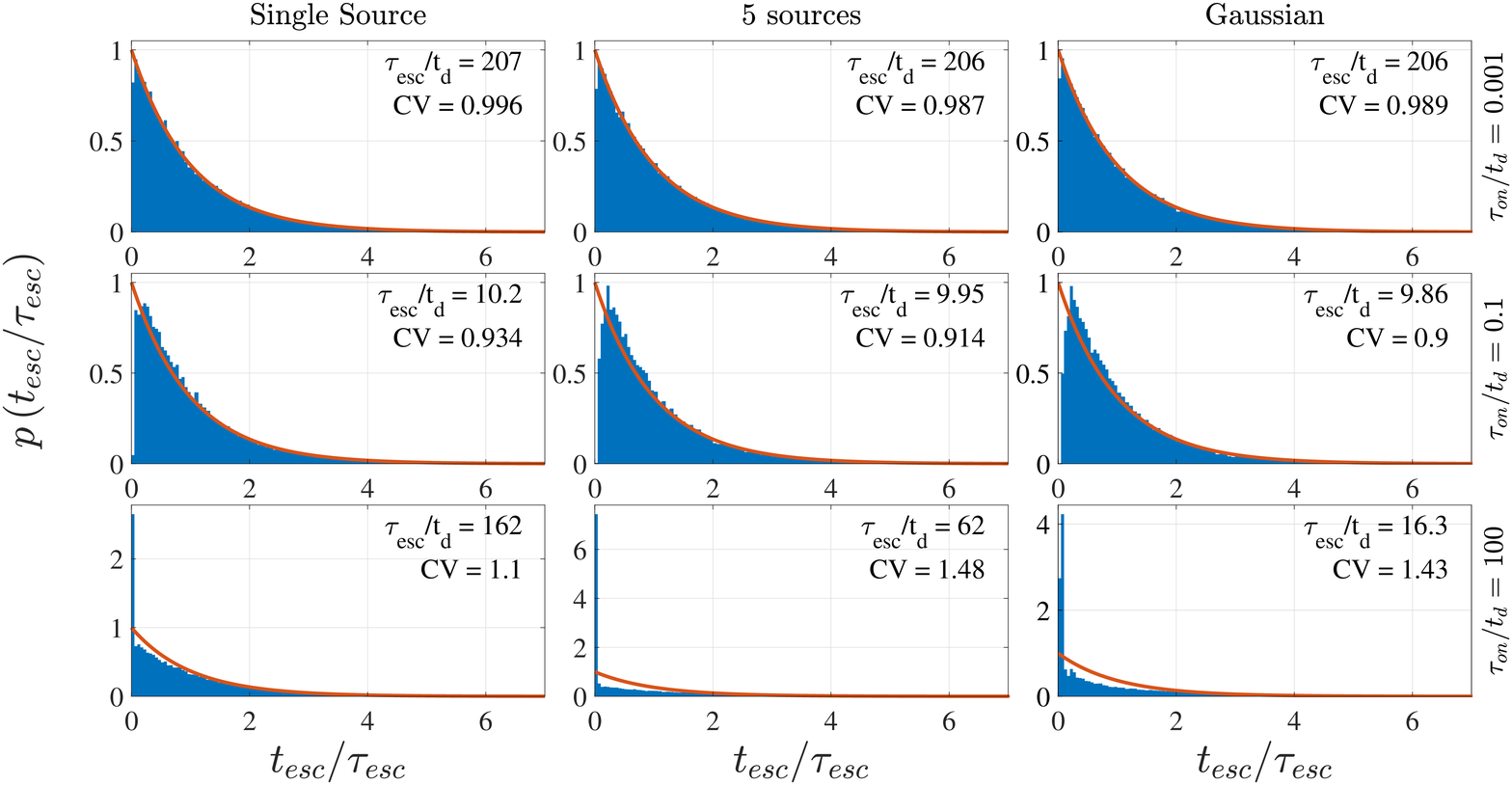}
\caption{\label{fig:meth}
Probability density functions of the escape time. The three rows correspond to the three indicated values of $\tau_{on}/t_d$ (0.001, 0.1 and 100). The different columns correspond to three different implementations of the active force with a single source, five sources and the Gaussian force, as indicated. In each panel, the individual escape time values $t_{esc}$ are scaled according to the mean escape time $\tau_{esc}$. For all the cases, we simulated the escape with the following parameters: $\alpha=5$, $\beta=10$ and $x_{esc}=x_d$.
}
\end{figure*}

The escape time is not fully characterized by the mean value because there is no guarantee that the PDF is exponential. The deviation from an exponential distribution may be characterized by the coefficient of variation, $CV\equiv\sigma_{esc}/\tau_{esc}$, where $\sigma_{esc}$ is the standard deviation of the escape time. In Fig. \ref{fig:meth}, we present the PDFs of the escape times for different parameters and the three implementations of the active force. In all cases, $\alpha=5$, $x_{esc}=x_d$, $\beta=10$, and$p_{on}=0.1$. For the shortest correlation time considered, $\tau_{on}=0.001t_d$, for all implementations, the mean escape time is similar, and the distribution is very close to exponential, $CV=0.989$\textendash$0.996$. The red lines present the exponential probability density with the same mean of the simulations. For the longer correlation time, $\tau_{on}=0.1t_d$, the deviation from exponential is apparent for all three active force implementations, and the probability density is narrower than exponential in the sense that the coefficient of variation is smaller than one, $CV=0.9$\textendash$0.934$. For the longest correlation time considered, $\tau_{on}=100t_d$, the probability density becomes broader than exponential for the active force implementations, $CV=1.1$\textendash$1.48$.

\subsection{Mean escape times from non-harmonic confining potentials}\label{ss:met_nh}

\begin{figure*}[b]
\includegraphics[width=\linewidth]{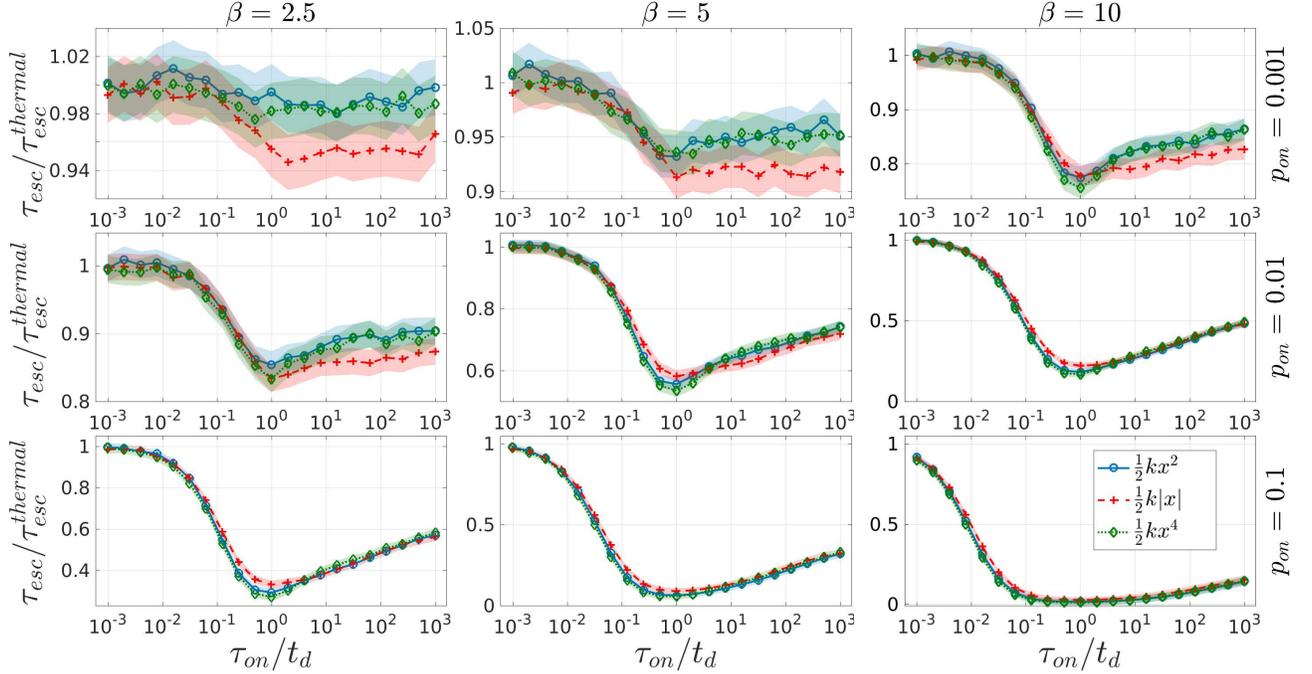}
\caption{\label{fig:pots}
Mean escape time vs. the correlation time of the active force for several trapping potentials. The three rows correspond to the three indicated values of $p_{on}$ (0.001, 0.01, and 0.1). The different columns correspond to three indicated values of $\beta (2.5, 5, and 10)$. The three lines in each panel correspond to three trapping potentials (harmonic, V-shaped and quartic). Each potential's mean escape times are normalized by the potential's thermal escape time ($225.2\pm1.1$, $120.6\pm0.6$ and $338.6\pm1.7$ $t_d$ for the harmonic, V-shaped, and quartic potentials, respectively). For all the cases, we simulated the escape with the following parameters: $\alpha=5$ and $x_{esc}=x_d$, using the Gaussian force implementation for the active force.
}
\end{figure*}

In order to investigate the relevance of our results to other confining potentials, we simulated the escape dynamics of particles subjected to thermal and active noise confined by two non-harmonic potentials. The first is the V-shaped potential, $U(x)=\alpha k_B T \cdot |x/x_d|$. The second potential is $U(x)=\alpha k_B T\cdot (x/x_d)^4$. The dynamics is described by eq. \eqref{eq:eomdim} with the term $-k\cdot x$ replaced by $-dU(x)/dx$.
For the V-shaped potential, the force at $x=0$ is defined to be zero. In order to compare the escape times from the different confining potentials, we set the energy barrier to be equal in all the potentials. We also used the same second moment and correlation time of the Gaussian active force. In Fig. \ref{fig:pots}, we present the normalized mean escape times (the mean escape time from each confining potential is normalized by the mean thermal escape time from the same potential) for different amplitudes of the active force as dictated by $\beta$ and $p_{on}$ values. The mean escape times are presented against the active force auto-correlation time. Note that the constant $p_{on}$ value in each panel implies that only the ratio between $\tau_{on}/\tau_{tot}$ is constant and not the cycle time.
We see that for all three potentials, the normalized mean escape time is very similar. Moreover, for all cases, the minimal mean escape time is obtained for the same auto-correlation time of the active force. These results suggest that our findings are not limited to the case of escape from harmonic potential, but they are likely to be valid for other confining potentials.

\section{Discussion}\label{sec:dis}

The inherent stochastic and non-equilibrium nature of active matter makes it difficult to find a universal description, similar to equilibrium statistical mechanics.
A common approach is to attempt to define an effective temperature that can describe some or all of the statistics of the out-of-equilibrium system.
However, the use of effective temperatures was shown to be limited for many systems. Moreover, the effective temperature is not uniquely defined.
Nevertheless, it is often tempting to stretch the use of the effective temperature beyond its range of validity.

The dynamics within and outside of a confining potential is important for many processes. For particles subjected to thermal noise alone, the celebrated Kramers' expression provides a relatively simple dependence of the mean escape time on the height of the energy barrier.

We found that for short to intermediate correlation times of the active force, an effective temperature combined with Kramers' expression well describes the mean escape time, while for long correlation times, the regular Kramers' expression, combined with modified energy barriers, provides an accurate description of the mean escape time.
In the following, we detail the limitations and range of the validity of these predictions.
An important point is that the particles considered are not overdamped (often the arguments in favor of considering the overdamped limit stem from the long range dynamics, and here, we focused on the dynamics within and outside of a single trap).

We derived analytically the second moments of the position and the velocity, which by analogy with the thermal case were used to define the potential energy and kinetic energy effective temperatures, $T_x$ and $T_v$, respectively. The expressions we derived using the direct integration in the time domain are somewhat different from previously derived expressions that were based on analysis in the frequency domain \cite{Benisaac2011,ben2015modeling,nandi2019erratum}. The effective temperatures depend on the second moment of the active force and its correlation time, regardless of the specific implementation of the active force (e.g., a single source or the Gaussian colored noise).
Importantly, the expressions provide the time dependence of the second moments and thereby the convergence time to the limiting value. Obviously, when the typical escape time from a trap is shorter than the convergence time to the saturation value of the second moments, one need not expect the effective temperatures based on the infinite time limit to be relevant for the description of the escape process. The convergence time depends on the friction and the confining potential characteristics but also on the active force correlation time. Therefore, correlation times of the active force that are longer than the time scale of interest render the effective temperatures useless. We found that in all cases, the simulated second moments of the position and the velocity of trapped particles were in excellent agreement with the calculated ones, regardless of the force implementation. Nevertheless, one should bear in mind that the effective temperatures depend on the confining potential and, therefore, are not similar to the thermal temperature, which is independent of the energy landscape.

We found that the probability density functions of the position and the velocity of trapped particles depend on the active force implementation. For the single source, the value of the force, while "on", is constant, which if remaining active for long enough periods, drives the particle to a stable point. For the Gaussian implementation, the amplitude of the active force changes continuously, and there is no stable point. Therefore, for the single source, the position’s PDF is tri-modal if the active force’s correlation time and amplitude are large enough, while for the Gaussian force, they are uni-modal. For all implementations of the active force, one may find a tri-modal distribution of the velocity for cases in which the confining potential is shallow, and the particle may reach a terminal velocity dictated by the friction and active force’s amplitude.

For particles that can escape from the confining potential, the PDFs are considerably different. The position PDF is truncated at the escape points, and this in turn, also affects the velocity PDF. The second moments of the position and the velocity are also affected by the escape, with the second moment of the position being always smaller due to the escape, as expected, while the second moment of the velocity can be smaller or greater (depending on the active force and potential parameters). Therefore, the relevance of the effective temperatures for the description of the escape process is not obvious.

We found that the mean escape time depends on the active force implementation, suggesting that the second moment and the correlation time are not sufficient for the full description of the escape process. We also showed that increasing the number of sources results in behavior similar to the Gaussian force implementation, as expected. For all implementations of the force, we found that for a fixed amplitude of the force, there is an optimal correlation time, which yields the minimal escape time. The appearance of the optimal correlation time is not limited to the harmonic potential, and we also found it for the V-shaped and polynomial potentials.
Using the potential effective temperature in Kramers' expression gives a good approximation for short and intermediate correlation times, and it also captures the minimal escape time. As expected, it fails for correlation times that are longer or of the same order of the mean escape time. For the longer correlation times, the effective temperatures that are based on the steady state probability density functions of trapped particles are not representative of the dynamics of particles that can escape because they escape before the active force statistics reaches its steady state. The kinetic energy effective temperature seems to be less relevant for the description of the escape process. It is important to note that we considered a system that is not overdamped, and this is the reason that the kinetic energy effective temperature is not useful in describing an escape at a fixed position. It is also worth noting that the difference between the potential and kinetic energy effective temperatures is larger for longer correlation times. The distribution of the escape time is not always exponential, and it varies with the active force and confining potential characteristics.

For long correlation times of the active force, it varies only slightly before the particle escapes. This allowed us to consider a modified confining potential and derive an analytical approximation for the mean escape time. The simulations showed an excellent agreement with this approximation.

Interestingly, we found that the normalized mean escape time is very similar for different confining potentials. In particular, the auto-correlation time of the active force for which the mean escape time is minimal was found to be almost identical for escape from all three confining potentials that we investigated. This in turn suggests that our results may not be limited to harmonic confining potential.

\section{Summary}\label{sec:sum}

To summarize, we investigated the dynamics of particles subjected to thermal and active noises within and outside of a confining potential. We derived analytical expressions for the effective temperatures and showed that the kinetic and potential energy effective temperatures are different. Numerical simulations showed that despite capturing the second moments of the position and the velocity, the probability density functions may show large deviations from Gaussian distributions even for trapped particles; therefore, the effective temperatures do not provide a full description of the statistical properties of the active particles. For the escape of the active particles, we found that Kramers' expression, with the potential energy effective temperature replacing the thermal temperature, gives a very good approximation for the mean escape time for short to intermediate auto-correlation times of the active force, but it fails for correlation times that are larger than or comparable to the mean escape time. An analytical expression, derived for the limit of long correlation times, showed excellent agreement with simulation results.

Our results are relevant to many systems in which an active force plays an important role in non-equilibrium dynamics. Such systems include living matter \cite{Marchetti2013} (biofilaments and molecular motors \cite{Son2017} in vitro or in vivo, collections of motile microorganisms \cite{Elgeti2015}, animal herds, chemical and mechanical imitations \cite{Marchetti2013}), active colloids \cite{Palacci2010,Ginot2015,Levis2015,Han2017}, mixtures of active particles \cite{Wittmann2018}, active particles in crowded environments \cite{Bechinger2016}, and active glasses \cite{Ni2013,nandi2017nonequilibrium,nandi2019erratum}. In spatially extended systems, the active escape process we explored here can be an individual event of a larger scale dynamics and the reorganization of the system. The interpolation between the short-intermediate correlation time and the long correlation time limits and the generalization to an arbitrary confining potential are left for future work.

%%%%
\begin{acknowledgments}
The authors are grateful to Samara T. Bel for proofreading the manuscript. 
NSG acknowledges support from the ISF (Grant No. 580/12).
\end{acknowledgments}
\bibliographystyle{apsrev4-2}
\bibliography{apssamp}% Produces the bibliography via BibTeX.

\providecommand{\noopsort}[1]{}\providecommand{\singleletter}[1]{#1}%
\begin{thebibliography}{56}%
\makeatletter
\providecommand \@ifxundefined [1]{%
 \@ifx{#1\undefined}
}%
\providecommand \@ifnum [1]{%
 \ifnum #1\expandafter \@firstoftwo
 \else \expandafter \@secondoftwo
 \fi
}%
\providecommand \@ifx [1]{%
 \ifx #1\expandafter \@firstoftwo
 \else \expandafter \@secondoftwo
 \fi
}%
\providecommand \natexlab [1]{#1}%
\providecommand \enquote  [1]{``#1''}%
\providecommand \bibnamefont  [1]{#1}%
\providecommand \bibfnamefont [1]{#1}%
\providecommand \citenamefont [1]{#1}%
\providecommand \href@noop [0]{\@secondoftwo}%
\providecommand \href [0]{\begingroup \@sanitize@url \@href}%
\providecommand \@href[1]{\@@startlink{#1}\@@href}%
\providecommand \@@href[1]{\endgroup#1\@@endlink}%
\providecommand \@sanitize@url [0]{\catcode `\\12\catcode `\$12\catcode
  `\&12\catcode `\#12\catcode `\^12\catcode `\_12\catcode `\%12\relax}%
\providecommand \@@startlink[1]{}%
\providecommand \@@endlink[0]{}%
\providecommand \url  [0]{\begingroup\@sanitize@url \@url }%
\providecommand \@url [1]{\endgroup\@href {#1}{\urlprefix }}%
\providecommand \urlprefix  [0]{URL }%
\providecommand \Eprint [0]{\href }%
\providecommand \doibase [0]{https://doi.org/}%
\providecommand \selectlanguage [0]{\@gobble}%
\providecommand \bibinfo  [0]{\@secondoftwo}%
\providecommand \bibfield  [0]{\@secondoftwo}%
\providecommand \translation [1]{[#1]}%
\providecommand \BibitemOpen [0]{}%
\providecommand \bibitemStop [0]{}%
\providecommand \bibitemNoStop [0]{.\EOS\space}%
\providecommand \EOS [0]{\spacefactor3000\relax}%
\providecommand \BibitemShut  [1]{\csname bibitem#1\endcsname}%
\let\auto@bib@innerbib\@empty
%</preamble>
\bibitem [{\citenamefont {Ramaswamy}(2010)}]{ramaswamy2010mechanics}%
  \BibitemOpen
  \bibfield  {author} {\bibinfo {author} {\bibfnamefont {S.}~\bibnamefont
  {Ramaswamy}},\ }\href@noop {} {\bibfield  {journal} {\bibinfo  {journal}
  {Annu. Rev. Condens. Matter Phys.}\ }\textbf {\bibinfo {volume} {1}},\
  \bibinfo {pages} {323} (\bibinfo {year} {2010})}\BibitemShut {NoStop}%
\bibitem [{\citenamefont {Marchetti}\ \emph
  {et~al.}(2013{\natexlab{a}})\citenamefont {Marchetti}, \citenamefont
  {Joanny}, \citenamefont {Ramaswamy}, \citenamefont {Liverpool}, \citenamefont
  {Prost}, \citenamefont {Rao},\ and\ \citenamefont
  {Simha}}]{marchetti2013hydrodynamics}%
  \BibitemOpen
  \bibfield  {author} {\bibinfo {author} {\bibfnamefont {M.~C.}\ \bibnamefont
  {Marchetti}}, \bibinfo {author} {\bibfnamefont {J.-F.}\ \bibnamefont
  {Joanny}}, \bibinfo {author} {\bibfnamefont {S.}~\bibnamefont {Ramaswamy}},
  \bibinfo {author} {\bibfnamefont {T.~B.}\ \bibnamefont {Liverpool}}, \bibinfo
  {author} {\bibfnamefont {J.}~\bibnamefont {Prost}}, \bibinfo {author}
  {\bibfnamefont {M.}~\bibnamefont {Rao}},\ and\ \bibinfo {author}
  {\bibfnamefont {R.~A.}\ \bibnamefont {Simha}},\ }\href@noop {} {\bibfield
  {journal} {\bibinfo  {journal} {Reviews of Modern Physics}\ }\textbf
  {\bibinfo {volume} {85}},\ \bibinfo {pages} {1143} (\bibinfo {year}
  {2013}{\natexlab{a}})}\BibitemShut {NoStop}%
\bibitem [{\citenamefont {Ramaswamy}(2017)}]{ramaswamy2017active}%
  \BibitemOpen
  \bibfield  {author} {\bibinfo {author} {\bibfnamefont {S.}~\bibnamefont
  {Ramaswamy}},\ }\href@noop {} {\bibfield  {journal} {\bibinfo  {journal}
  {Journal of Statistical Mechanics: Theory and Experiment}\ }\textbf {\bibinfo
  {volume} {2017}},\ \bibinfo {pages} {054002} (\bibinfo {year}
  {2017})}\BibitemShut {NoStop}%
\bibitem [{\citenamefont {Bernheim-Groswasser}\ \emph
  {et~al.}(2018)\citenamefont {Bernheim-Groswasser}, \citenamefont {Gov},
  \citenamefont {Safran},\ and\ \citenamefont {Tzlil}}]{Bernheim2018}%
  \BibitemOpen
  \bibfield  {author} {\bibinfo {author} {\bibfnamefont {A.}~\bibnamefont
  {Bernheim-Groswasser}}, \bibinfo {author} {\bibfnamefont {N.~S.}\
  \bibnamefont {Gov}}, \bibinfo {author} {\bibfnamefont {S.~A.}\ \bibnamefont
  {Safran}},\ and\ \bibinfo {author} {\bibfnamefont {S.}~\bibnamefont
  {Tzlil}},\ }\href@noop {} {\bibfield  {journal} {\bibinfo  {journal}
  {Advanced Materials}\ }\textbf {\bibinfo {volume} {30}},\ \bibinfo {pages}
  {1707028} (\bibinfo {year} {2018})}\BibitemShut {NoStop}%
\bibitem [{\citenamefont {Prost}\ \emph {et~al.}(2015)\citenamefont {Prost},
  \citenamefont {J{\"u}licher},\ and\ \citenamefont
  {Joanny}}]{prost2015active}%
  \BibitemOpen
  \bibfield  {author} {\bibinfo {author} {\bibfnamefont {J.}~\bibnamefont
  {Prost}}, \bibinfo {author} {\bibfnamefont {F.}~\bibnamefont
  {J{\"u}licher}},\ and\ \bibinfo {author} {\bibfnamefont {J.-F.}\ \bibnamefont
  {Joanny}},\ }\href@noop {} {\bibfield  {journal} {\bibinfo  {journal} {Nature
  Physics}\ }\textbf {\bibinfo {volume} {11}},\ \bibinfo {pages} {111}
  (\bibinfo {year} {2015})}\BibitemShut {NoStop}%
\bibitem [{\citenamefont {Dreyfus}\ \emph {et~al.}(2005)\citenamefont
  {Dreyfus}, \citenamefont {Baudry}, \citenamefont {Roper}, \citenamefont
  {Fermigier}, \citenamefont {Stone},\ and\ \citenamefont
  {Bibette}}]{dreyfus2005microscopic}%
  \BibitemOpen
  \bibfield  {author} {\bibinfo {author} {\bibfnamefont {R.}~\bibnamefont
  {Dreyfus}}, \bibinfo {author} {\bibfnamefont {J.}~\bibnamefont {Baudry}},
  \bibinfo {author} {\bibfnamefont {M.~L.}\ \bibnamefont {Roper}}, \bibinfo
  {author} {\bibfnamefont {M.}~\bibnamefont {Fermigier}}, \bibinfo {author}
  {\bibfnamefont {H.~A.}\ \bibnamefont {Stone}},\ and\ \bibinfo {author}
  {\bibfnamefont {J.}~\bibnamefont {Bibette}},\ }\href@noop {} {\bibfield
  {journal} {\bibinfo  {journal} {Nature}\ }\textbf {\bibinfo {volume} {437}},\
  \bibinfo {pages} {862} (\bibinfo {year} {2005})}\BibitemShut {NoStop}%
\bibitem [{\citenamefont {Palacci}\ \emph {et~al.}(2013)\citenamefont
  {Palacci}, \citenamefont {Sacanna}, \citenamefont {Steinberg}, \citenamefont
  {Pine},\ and\ \citenamefont {Chaikin}}]{palacci2013living}%
  \BibitemOpen
  \bibfield  {author} {\bibinfo {author} {\bibfnamefont {J.}~\bibnamefont
  {Palacci}}, \bibinfo {author} {\bibfnamefont {S.}~\bibnamefont {Sacanna}},
  \bibinfo {author} {\bibfnamefont {A.~P.}\ \bibnamefont {Steinberg}}, \bibinfo
  {author} {\bibfnamefont {D.~J.}\ \bibnamefont {Pine}},\ and\ \bibinfo
  {author} {\bibfnamefont {P.~M.}\ \bibnamefont {Chaikin}},\ }\href@noop {}
  {\bibfield  {journal} {\bibinfo  {journal} {Science}\ }\textbf {\bibinfo
  {volume} {339}},\ \bibinfo {pages} {936} (\bibinfo {year}
  {2013})}\BibitemShut {NoStop}%
\bibitem [{\citenamefont {Cates}\ and\ \citenamefont
  {Tailleur}(2015)}]{cates2015motility}%
  \BibitemOpen
  \bibfield  {author} {\bibinfo {author} {\bibfnamefont {M.~E.}\ \bibnamefont
  {Cates}}\ and\ \bibinfo {author} {\bibfnamefont {J.}~\bibnamefont
  {Tailleur}},\ }\href@noop {} {\bibfield  {journal} {\bibinfo  {journal}
  {Annu. Rev. Condens. Matter Phys.}\ }\textbf {\bibinfo {volume} {6}},\
  \bibinfo {pages} {219} (\bibinfo {year} {2015})}\BibitemShut {NoStop}%
\bibitem [{\citenamefont {Z{\"o}ttl}\ and\ \citenamefont
  {Stark}(2016)}]{zottl2016emergent}%
  \BibitemOpen
  \bibfield  {author} {\bibinfo {author} {\bibfnamefont {A.}~\bibnamefont
  {Z{\"o}ttl}}\ and\ \bibinfo {author} {\bibfnamefont {H.}~\bibnamefont
  {Stark}},\ }\href@noop {} {\bibfield  {journal} {\bibinfo  {journal} {Journal
  of Physics: Condensed Matter}\ }\textbf {\bibinfo {volume} {28}},\ \bibinfo
  {pages} {253001} (\bibinfo {year} {2016})}\BibitemShut {NoStop}%
\bibitem [{\citenamefont {Berthier}\ and\ \citenamefont
  {Kurchan}(2013)}]{berthier2013non}%
  \BibitemOpen
  \bibfield  {author} {\bibinfo {author} {\bibfnamefont {L.}~\bibnamefont
  {Berthier}}\ and\ \bibinfo {author} {\bibfnamefont {J.}~\bibnamefont
  {Kurchan}},\ }\href@noop {} {\bibfield  {journal} {\bibinfo  {journal}
  {Nature Physics}\ }\textbf {\bibinfo {volume} {9}},\ \bibinfo {pages} {310}
  (\bibinfo {year} {2013})}\BibitemShut {NoStop}%
\bibitem [{\citenamefont {{Berthier}}\ \emph {et~al.}(2019)\citenamefont
  {{Berthier}}, \citenamefont {{Flenner}},\ and\ \citenamefont
  {{Szamel}}}]{Berthier2019}%
  \BibitemOpen
  \bibfield  {author} {\bibinfo {author} {\bibfnamefont {L.}~\bibnamefont
  {{Berthier}}}, \bibinfo {author} {\bibfnamefont {E.}~\bibnamefont
  {{Flenner}}},\ and\ \bibinfo {author} {\bibfnamefont {G.}~\bibnamefont
  {{Szamel}}},\ }\href@noop {} {\bibfield  {journal} {\bibinfo  {journal}
  {arXiv e-prints}\ ,\ \bibinfo {eid} {arXiv:1902.08580}} (\bibinfo {year}
  {2019})},\ \Eprint {https://arxiv.org/abs/1902.08580} {arXiv:1902.08580
  [cond-mat.soft]} \BibitemShut {NoStop}%
\bibitem [{\citenamefont {{Janssen}}(2019)}]{Janssen2019}%
  \BibitemOpen
  \bibfield  {author} {\bibinfo {author} {\bibfnamefont {L.~M.~C.}\
  \bibnamefont {{Janssen}}},\ }\href@noop {} {\bibfield  {journal} {\bibinfo
  {journal} {arXiv e-prints}\ ,\ \bibinfo {eid} {arXiv:1906.03678}} (\bibinfo
  {year} {2019})},\ \Eprint {https://arxiv.org/abs/1906.03678}
  {arXiv:1906.03678 [cond-mat.soft]} \BibitemShut {NoStop}%
\bibitem [{\citenamefont {Henkes}\ \emph {et~al.}(2011)\citenamefont {Henkes},
  \citenamefont {Fily},\ and\ \citenamefont {Marchetti}}]{henkes2011active}%
  \BibitemOpen
  \bibfield  {author} {\bibinfo {author} {\bibfnamefont {S.}~\bibnamefont
  {Henkes}}, \bibinfo {author} {\bibfnamefont {Y.}~\bibnamefont {Fily}},\ and\
  \bibinfo {author} {\bibfnamefont {M.~C.}\ \bibnamefont {Marchetti}},\
  }\href@noop {} {\bibfield  {journal} {\bibinfo  {journal} {Physical Review
  E}\ }\textbf {\bibinfo {volume} {84}},\ \bibinfo {pages} {040301} (\bibinfo
  {year} {2011})}\BibitemShut {NoStop}%
\bibitem [{\citenamefont {Berthier}(2014)}]{berthier2014nonequilibrium}%
  \BibitemOpen
  \bibfield  {author} {\bibinfo {author} {\bibfnamefont {L.}~\bibnamefont
  {Berthier}},\ }\href@noop {} {\bibfield  {journal} {\bibinfo  {journal}
  {Physical review letters}\ }\textbf {\bibinfo {volume} {112}},\ \bibinfo
  {pages} {220602} (\bibinfo {year} {2014})}\BibitemShut {NoStop}%
\bibitem [{\citenamefont {Bi}\ \emph {et~al.}(2014)\citenamefont {Bi},
  \citenamefont {Lopez}, \citenamefont {Schwarz},\ and\ \citenamefont
  {Manning}}]{bi2014energy}%
  \BibitemOpen
  \bibfield  {author} {\bibinfo {author} {\bibfnamefont {D.}~\bibnamefont
  {Bi}}, \bibinfo {author} {\bibfnamefont {J.~H.}\ \bibnamefont {Lopez}},
  \bibinfo {author} {\bibfnamefont {J.}~\bibnamefont {Schwarz}},\ and\ \bibinfo
  {author} {\bibfnamefont {M.~L.}\ \bibnamefont {Manning}},\ }\href@noop {}
  {\bibfield  {journal} {\bibinfo  {journal} {Soft matter}\ }\textbf {\bibinfo
  {volume} {10}},\ \bibinfo {pages} {1885} (\bibinfo {year}
  {2014})}\BibitemShut {NoStop}%
\bibitem [{\citenamefont {Garcia}\ \emph {et~al.}(2015)\citenamefont {Garcia},
  \citenamefont {Hannezo}, \citenamefont {Elgeti}, \citenamefont {Joanny},
  \citenamefont {Silberzan},\ and\ \citenamefont {Gov}}]{garcia2015physics}%
  \BibitemOpen
  \bibfield  {author} {\bibinfo {author} {\bibfnamefont {S.}~\bibnamefont
  {Garcia}}, \bibinfo {author} {\bibfnamefont {E.}~\bibnamefont {Hannezo}},
  \bibinfo {author} {\bibfnamefont {J.}~\bibnamefont {Elgeti}}, \bibinfo
  {author} {\bibfnamefont {J.-F.}\ \bibnamefont {Joanny}}, \bibinfo {author}
  {\bibfnamefont {P.}~\bibnamefont {Silberzan}},\ and\ \bibinfo {author}
  {\bibfnamefont {N.~S.}\ \bibnamefont {Gov}},\ }\href@noop {} {\bibfield
  {journal} {\bibinfo  {journal} {Proceedings of the National Academy of
  Sciences}\ }\textbf {\bibinfo {volume} {112}},\ \bibinfo {pages} {15314}
  (\bibinfo {year} {2015})}\BibitemShut {NoStop}%
\bibitem [{\citenamefont {Bi}\ \emph {et~al.}(2016)\citenamefont {Bi},
  \citenamefont {Yang}, \citenamefont {Marchetti},\ and\ \citenamefont
  {Manning}}]{bi2016motility}%
  \BibitemOpen
  \bibfield  {author} {\bibinfo {author} {\bibfnamefont {D.}~\bibnamefont
  {Bi}}, \bibinfo {author} {\bibfnamefont {X.}~\bibnamefont {Yang}}, \bibinfo
  {author} {\bibfnamefont {M.~C.}\ \bibnamefont {Marchetti}},\ and\ \bibinfo
  {author} {\bibfnamefont {M.~L.}\ \bibnamefont {Manning}},\ }\href@noop {}
  {\bibfield  {journal} {\bibinfo  {journal} {Physical Review X}\ }\textbf
  {\bibinfo {volume} {6}},\ \bibinfo {pages} {021011} (\bibinfo {year}
  {2016})}\BibitemShut {NoStop}%
\bibitem [{\citenamefont {Cugliandolo}(2011)}]{cugliandolo2011effective}%
  \BibitemOpen
  \bibfield  {author} {\bibinfo {author} {\bibfnamefont {L.~F.}\ \bibnamefont
  {Cugliandolo}},\ }\href@noop {} {\bibfield  {journal} {\bibinfo  {journal}
  {Journal of Physics A: Mathematical and Theoretical}\ }\textbf {\bibinfo
  {volume} {44}},\ \bibinfo {pages} {483001} (\bibinfo {year}
  {2011})}\BibitemShut {NoStop}%
\bibitem [{\citenamefont {Wang}\ and\ \citenamefont
  {Wolynes}(2011)}]{wang2011}%
  \BibitemOpen
  \bibfield  {author} {\bibinfo {author} {\bibfnamefont {S.}~\bibnamefont
  {Wang}}\ and\ \bibinfo {author} {\bibfnamefont {P.~G.}\ \bibnamefont
  {Wolynes}},\ }\href@noop {} {\bibfield  {journal} {\bibinfo  {journal} {J.
  Chem. Phys.}\ }\textbf {\bibinfo {volume} {135}},\ \bibinfo {pages} {051101}
  (\bibinfo {year} {2011})}\BibitemShut {NoStop}%
\bibitem [{\citenamefont {Puglisi}\ \emph {et~al.}(2017)\citenamefont
  {Puglisi}, \citenamefont {Sarracino},\ and\ \citenamefont
  {Vulpiani}}]{Puglisi2017}%
  \BibitemOpen
  \bibfield  {author} {\bibinfo {author} {\bibfnamefont {A.}~\bibnamefont
  {Puglisi}}, \bibinfo {author} {\bibfnamefont {A.}~\bibnamefont {Sarracino}},\
  and\ \bibinfo {author} {\bibfnamefont {A.}~\bibnamefont {Vulpiani}},\ }\href
  {https://doi.org/https://doi.org/10.1016/j.physrep.2017.09.001} {\bibfield
  {journal} {\bibinfo  {journal} {Physics Reports}\ }\textbf {\bibinfo {volume}
  {709-710}},\ \bibinfo {pages} {1 } (\bibinfo {year} {2017})},\ \bibinfo
  {note} {temperature in and out of equilibrium: a review of concepts, tools
  and attempts}\BibitemShut {NoStop}%
\bibitem [{\citenamefont {Betz}\ \emph {et~al.}(2009)\citenamefont {Betz},
  \citenamefont {Lenz}, \citenamefont {Joanny},\ and\ \citenamefont
  {Sykes}}]{Betz2009}%
  \BibitemOpen
  \bibfield  {author} {\bibinfo {author} {\bibfnamefont {T.}~\bibnamefont
  {Betz}}, \bibinfo {author} {\bibfnamefont {M.}~\bibnamefont {Lenz}}, \bibinfo
  {author} {\bibfnamefont {J.-F.}\ \bibnamefont {Joanny}},\ and\ \bibinfo
  {author} {\bibfnamefont {C.}~\bibnamefont {Sykes}},\ }\href@noop {}
  {\bibfield  {journal} {\bibinfo  {journal} {Proceedings of the National
  Academy of Sciences}\ }\textbf {\bibinfo {volume} {106}},\ \bibinfo {pages}
  {15320} (\bibinfo {year} {2009})}\BibitemShut {NoStop}%
\bibitem [{\citenamefont {Park}\ \emph {et~al.}(2010)\citenamefont {Park},
  \citenamefont {Best}, \citenamefont {Auth}, \citenamefont {Gov},
  \citenamefont {Safran}, \citenamefont {Popescu}, \citenamefont {Suresh},\
  and\ \citenamefont {Feld}}]{Park2010}%
  \BibitemOpen
  \bibfield  {author} {\bibinfo {author} {\bibfnamefont {Y.}~\bibnamefont
  {Park}}, \bibinfo {author} {\bibfnamefont {C.~A.}\ \bibnamefont {Best}},
  \bibinfo {author} {\bibfnamefont {T.}~\bibnamefont {Auth}}, \bibinfo {author}
  {\bibfnamefont {N.~S.}\ \bibnamefont {Gov}}, \bibinfo {author} {\bibfnamefont
  {S.~A.}\ \bibnamefont {Safran}}, \bibinfo {author} {\bibfnamefont
  {G.}~\bibnamefont {Popescu}}, \bibinfo {author} {\bibfnamefont
  {S.}~\bibnamefont {Suresh}},\ and\ \bibinfo {author} {\bibfnamefont {M.~S.}\
  \bibnamefont {Feld}},\ }\href@noop {} {\bibfield  {journal} {\bibinfo
  {journal} {Proceedings of the National Academy of Sciences}\ }\textbf
  {\bibinfo {volume} {107}},\ \bibinfo {pages} {1289} (\bibinfo {year}
  {2010})}\BibitemShut {NoStop}%
\bibitem [{\citenamefont {Ben-Isaac}\ \emph {et~al.}(2011)\citenamefont
  {Ben-Isaac}, \citenamefont {Park}, \citenamefont {Popescu}, \citenamefont
  {Brown}, \citenamefont {Gov},\ and\ \citenamefont {Shokef}}]{Benisaac2011}%
  \BibitemOpen
  \bibfield  {author} {\bibinfo {author} {\bibfnamefont {E.}~\bibnamefont
  {Ben-Isaac}}, \bibinfo {author} {\bibfnamefont {Y.}~\bibnamefont {Park}},
  \bibinfo {author} {\bibfnamefont {G.}~\bibnamefont {Popescu}}, \bibinfo
  {author} {\bibfnamefont {F.~L.~H.}\ \bibnamefont {Brown}}, \bibinfo {author}
  {\bibfnamefont {N.~S.}\ \bibnamefont {Gov}},\ and\ \bibinfo {author}
  {\bibfnamefont {Y.}~\bibnamefont {Shokef}},\ }\href
  {https://doi.org/10.1103/PhysRevLett.106.238103} {\bibfield  {journal}
  {\bibinfo  {journal} {Phys. Rev. Lett.}\ }\textbf {\bibinfo {volume} {106}},\
  \bibinfo {pages} {238103} (\bibinfo {year} {2011})}\BibitemShut {NoStop}%
\bibitem [{\citenamefont {Berthier}\ and\ \citenamefont
  {Barrat}(2002)}]{Berthier2002}%
  \BibitemOpen
  \bibfield  {author} {\bibinfo {author} {\bibfnamefont {L.}~\bibnamefont
  {Berthier}}\ and\ \bibinfo {author} {\bibfnamefont {J.-L.}\ \bibnamefont
  {Barrat}},\ }\href@noop {} {\bibfield  {journal} {\bibinfo  {journal}
  {Physical review letters}\ }\textbf {\bibinfo {volume} {89}},\ \bibinfo
  {pages} {095702} (\bibinfo {year} {2002})}\BibitemShut {NoStop}%
\bibitem [{\citenamefont {O’Hern}\ \emph {et~al.}(2004)\citenamefont
  {O’Hern}, \citenamefont {Liu},\ and\ \citenamefont {Nagel}}]{Ohren2004}%
  \BibitemOpen
  \bibfield  {author} {\bibinfo {author} {\bibfnamefont {C.~S.}\ \bibnamefont
  {O’Hern}}, \bibinfo {author} {\bibfnamefont {A.~J.}\ \bibnamefont {Liu}},\
  and\ \bibinfo {author} {\bibfnamefont {S.~R.}\ \bibnamefont {Nagel}},\
  }\href@noop {} {\bibfield  {journal} {\bibinfo  {journal} {Physical review
  letters}\ }\textbf {\bibinfo {volume} {93}},\ \bibinfo {pages} {165702}
  (\bibinfo {year} {2004})}\BibitemShut {NoStop}%
\bibitem [{\citenamefont {Nandi}\ and\ \citenamefont
  {Gov}(2019)}]{nandi2019erratum}%
  \BibitemOpen
  \bibfield  {author} {\bibinfo {author} {\bibfnamefont {S.~K.}\ \bibnamefont
  {Nandi}}\ and\ \bibinfo {author} {\bibfnamefont {N.~S.}\ \bibnamefont
  {Gov}},\ }\href@noop {} {\bibfield  {journal} {\bibinfo  {journal} {The
  European Physical Journal E}\ }\textbf {\bibinfo {volume} {42}},\ \bibinfo
  {pages} {16} (\bibinfo {year} {2019})}\BibitemShut {NoStop}%
\bibitem [{\citenamefont {Nandi}\ and\ \citenamefont
  {Gov}(2017)}]{nandi2017nonequilibrium}%
  \BibitemOpen
  \bibfield  {author} {\bibinfo {author} {\bibfnamefont {S.~K.}\ \bibnamefont
  {Nandi}}\ and\ \bibinfo {author} {\bibfnamefont {N.~S.}\ \bibnamefont
  {Gov}},\ }\href@noop {} {\bibfield  {journal} {\bibinfo  {journal} {Soft
  matter}\ }\textbf {\bibinfo {volume} {13}},\ \bibinfo {pages} {7609}
  (\bibinfo {year} {2017})}\BibitemShut {NoStop}%
\bibitem [{\citenamefont {Nandi}\ \emph {et~al.}(2018)\citenamefont {Nandi},
  \citenamefont {Mandal}, \citenamefont {Bhuyan}, \citenamefont {Dasgupta},
  \citenamefont {Rao},\ and\ \citenamefont {Gov}}]{nandi2018random}%
  \BibitemOpen
  \bibfield  {author} {\bibinfo {author} {\bibfnamefont {S.~K.}\ \bibnamefont
  {Nandi}}, \bibinfo {author} {\bibfnamefont {R.}~\bibnamefont {Mandal}},
  \bibinfo {author} {\bibfnamefont {P.~J.}\ \bibnamefont {Bhuyan}}, \bibinfo
  {author} {\bibfnamefont {C.}~\bibnamefont {Dasgupta}}, \bibinfo {author}
  {\bibfnamefont {M.}~\bibnamefont {Rao}},\ and\ \bibinfo {author}
  {\bibfnamefont {N.~S.}\ \bibnamefont {Gov}},\ }\href@noop {} {\bibfield
  {journal} {\bibinfo  {journal} {Proceedings of the National Academy of
  Sciences}\ }\textbf {\bibinfo {volume} {115}},\ \bibinfo {pages} {7688}
  (\bibinfo {year} {2018})}\BibitemShut {NoStop}%
\bibitem [{\citenamefont {Cugliandolo}\ \emph {et~al.}(2019)\citenamefont
  {Cugliandolo}, \citenamefont {Gonnella},\ and\ \citenamefont
  {Petrelli}}]{cugliandolo2019effective}%
  \BibitemOpen
  \bibfield  {author} {\bibinfo {author} {\bibfnamefont {L.~F.}\ \bibnamefont
  {Cugliandolo}}, \bibinfo {author} {\bibfnamefont {G.}~\bibnamefont
  {Gonnella}},\ and\ \bibinfo {author} {\bibfnamefont {I.}~\bibnamefont
  {Petrelli}},\ }\href@noop {} {\bibfield  {journal} {\bibinfo  {journal}
  {Fluctuation and Noise Letters}\ ,\ \bibinfo {pages} {1940008}} (\bibinfo
  {year} {2019})}\BibitemShut {NoStop}%
\bibitem [{\citenamefont {Caprini}\ \emph {et~al.}(2019)\citenamefont
  {Caprini}, \citenamefont {Marini Bettolo~Marconi}, \citenamefont {Puglisi},\
  and\ \citenamefont {Vulpiani}}]{caprini2019active}%
  \BibitemOpen
  \bibfield  {author} {\bibinfo {author} {\bibfnamefont {L.}~\bibnamefont
  {Caprini}}, \bibinfo {author} {\bibfnamefont {U.}~\bibnamefont {Marini
  Bettolo~Marconi}}, \bibinfo {author} {\bibfnamefont {A.}~\bibnamefont
  {Puglisi}},\ and\ \bibinfo {author} {\bibfnamefont {A.}~\bibnamefont
  {Vulpiani}},\ }\href@noop {} {\bibfield  {journal} {\bibinfo  {journal} {The
  Journal of chemical physics}\ }\textbf {\bibinfo {volume} {150}},\ \bibinfo
  {pages} {024902} (\bibinfo {year} {2019})}\BibitemShut {NoStop}%
\bibitem [{\citenamefont {Woillez}\ \emph {et~al.}(2019)\citenamefont
  {Woillez}, \citenamefont {Zhao}, \citenamefont {Kafri}, \citenamefont
  {Lecomte},\ and\ \citenamefont {Tailleur}}]{Woillez2019}%
  \BibitemOpen
  \bibfield  {author} {\bibinfo {author} {\bibfnamefont {E.}~\bibnamefont
  {Woillez}}, \bibinfo {author} {\bibfnamefont {Y.}~\bibnamefont {Zhao}},
  \bibinfo {author} {\bibfnamefont {Y.}~\bibnamefont {Kafri}}, \bibinfo
  {author} {\bibfnamefont {V.}~\bibnamefont {Lecomte}},\ and\ \bibinfo {author}
  {\bibfnamefont {J.}~\bibnamefont {Tailleur}},\ }\href
  {https://doi.org/10.1103/PhysRevLett.122.258001} {\bibfield  {journal}
  {\bibinfo  {journal} {Phys. Rev. Lett.}\ }\textbf {\bibinfo {volume} {122}},\
  \bibinfo {pages} {258001} (\bibinfo {year} {2019})}\BibitemShut {NoStop}%
\bibitem [{\citenamefont {Fodor}\ \emph
  {et~al.}(2016{\natexlab{a}})\citenamefont {Fodor}, \citenamefont {Hayakawa},
  \citenamefont {Visco},\ and\ \citenamefont {van Wijland}}]{fodor2016active}%
  \BibitemOpen
  \bibfield  {author} {\bibinfo {author} {\bibfnamefont {{\'E}.}~\bibnamefont
  {Fodor}}, \bibinfo {author} {\bibfnamefont {H.}~\bibnamefont {Hayakawa}},
  \bibinfo {author} {\bibfnamefont {P.}~\bibnamefont {Visco}},\ and\ \bibinfo
  {author} {\bibfnamefont {F.}~\bibnamefont {van Wijland}},\ }\href@noop {}
  {\bibfield  {journal} {\bibinfo  {journal} {Physical Review E}\ }\textbf
  {\bibinfo {volume} {94}},\ \bibinfo {pages} {012610} (\bibinfo {year}
  {2016}{\natexlab{a}})}\BibitemShut {NoStop}%
\bibitem [{\citenamefont {Ben-Isaac}\ \emph {et~al.}(2015)\citenamefont
  {Ben-Isaac}, \citenamefont {Fodor}, \citenamefont {Visco}, \citenamefont
  {Van~Wijland},\ and\ \citenamefont {Gov}}]{ben2015modeling}%
  \BibitemOpen
  \bibfield  {author} {\bibinfo {author} {\bibfnamefont {E.}~\bibnamefont
  {Ben-Isaac}}, \bibinfo {author} {\bibfnamefont {{\'E}.}~\bibnamefont
  {Fodor}}, \bibinfo {author} {\bibfnamefont {P.}~\bibnamefont {Visco}},
  \bibinfo {author} {\bibfnamefont {F.}~\bibnamefont {Van~Wijland}},\ and\
  \bibinfo {author} {\bibfnamefont {N.~S.}\ \bibnamefont {Gov}},\ }\href@noop
  {} {\bibfield  {journal} {\bibinfo  {journal} {Physical Review E}\ }\textbf
  {\bibinfo {volume} {92}},\ \bibinfo {pages} {012716} (\bibinfo {year}
  {2015})}\BibitemShut {NoStop}%
\bibitem [{\citenamefont {Razin}\ \emph {et~al.}(2019)\citenamefont {Razin},
  \citenamefont {Voituriez},\ and\ \citenamefont {Gov}}]{razin2019signatures}%
  \BibitemOpen
  \bibfield  {author} {\bibinfo {author} {\bibfnamefont {N.}~\bibnamefont
  {Razin}}, \bibinfo {author} {\bibfnamefont {R.}~\bibnamefont {Voituriez}},\
  and\ \bibinfo {author} {\bibfnamefont {N.~S.}\ \bibnamefont {Gov}},\
  }\href@noop {} {\bibfield  {journal} {\bibinfo  {journal} {Physical Review
  E}\ }\textbf {\bibinfo {volume} {99}},\ \bibinfo {pages} {022419} (\bibinfo
  {year} {2019})}\BibitemShut {NoStop}%
\bibitem [{\citenamefont {Fodor}\ \emph {et~al.}(2015)\citenamefont {Fodor},
  \citenamefont {Guo}, \citenamefont {Gov}, \citenamefont {Visco},
  \citenamefont {Weitz},\ and\ \citenamefont {van
  Wijland}}]{fodor2015activity}%
  \BibitemOpen
  \bibfield  {author} {\bibinfo {author} {\bibfnamefont {{\'E}.}~\bibnamefont
  {Fodor}}, \bibinfo {author} {\bibfnamefont {M.}~\bibnamefont {Guo}}, \bibinfo
  {author} {\bibfnamefont {N.}~\bibnamefont {Gov}}, \bibinfo {author}
  {\bibfnamefont {P.}~\bibnamefont {Visco}}, \bibinfo {author} {\bibfnamefont
  {D.}~\bibnamefont {Weitz}},\ and\ \bibinfo {author} {\bibfnamefont
  {F.}~\bibnamefont {van Wijland}},\ }\href@noop {} {\bibfield  {journal}
  {\bibinfo  {journal} {EPL (Europhysics Letters)}\ }\textbf {\bibinfo {volume}
  {110}},\ \bibinfo {pages} {48005} (\bibinfo {year} {2015})}\BibitemShut
  {NoStop}%
\bibitem [{\citenamefont {Fodor}\ \emph
  {et~al.}(2016{\natexlab{b}})\citenamefont {Fodor}, \citenamefont {Ahmed},
  \citenamefont {Almonacid}, \citenamefont {Bussonnier}, \citenamefont {Gov},
  \citenamefont {Verlhac}, \citenamefont {Betz}, \citenamefont {Visco},\ and\
  \citenamefont {van Wijland}}]{fodor2016nonequilibrium}%
  \BibitemOpen
  \bibfield  {author} {\bibinfo {author} {\bibfnamefont {{\'E}.}~\bibnamefont
  {Fodor}}, \bibinfo {author} {\bibfnamefont {W.~W.}\ \bibnamefont {Ahmed}},
  \bibinfo {author} {\bibfnamefont {M.}~\bibnamefont {Almonacid}}, \bibinfo
  {author} {\bibfnamefont {M.}~\bibnamefont {Bussonnier}}, \bibinfo {author}
  {\bibfnamefont {N.~S.}\ \bibnamefont {Gov}}, \bibinfo {author} {\bibfnamefont
  {M.-H.}\ \bibnamefont {Verlhac}}, \bibinfo {author} {\bibfnamefont
  {T.}~\bibnamefont {Betz}}, \bibinfo {author} {\bibfnamefont {P.}~\bibnamefont
  {Visco}},\ and\ \bibinfo {author} {\bibfnamefont {F.}~\bibnamefont {van
  Wijland}},\ }\href@noop {} {\bibfield  {journal} {\bibinfo  {journal} {EPL
  (Europhysics Letters)}\ }\textbf {\bibinfo {volume} {116}},\ \bibinfo {pages}
  {30008} (\bibinfo {year} {2016}{\natexlab{b}})}\BibitemShut {NoStop}%
\bibitem [{\citenamefont {Ahmed}\ \emph {et~al.}(2018)\citenamefont {Ahmed},
  \citenamefont {Fodor}, \citenamefont {Almonacid}, \citenamefont {Bussonnier},
  \citenamefont {Verlhac}, \citenamefont {Gov}, \citenamefont {Visco},
  \citenamefont {van Wijland},\ and\ \citenamefont {Betz}}]{ahmed2018active}%
  \BibitemOpen
  \bibfield  {author} {\bibinfo {author} {\bibfnamefont {W.~W.}\ \bibnamefont
  {Ahmed}}, \bibinfo {author} {\bibfnamefont {E.}~\bibnamefont {Fodor}},
  \bibinfo {author} {\bibfnamefont {M.}~\bibnamefont {Almonacid}}, \bibinfo
  {author} {\bibfnamefont {M.}~\bibnamefont {Bussonnier}}, \bibinfo {author}
  {\bibfnamefont {M.-H.}\ \bibnamefont {Verlhac}}, \bibinfo {author}
  {\bibfnamefont {N.}~\bibnamefont {Gov}}, \bibinfo {author} {\bibfnamefont
  {P.}~\bibnamefont {Visco}}, \bibinfo {author} {\bibfnamefont
  {F.}~\bibnamefont {van Wijland}},\ and\ \bibinfo {author} {\bibfnamefont
  {T.}~\bibnamefont {Betz}},\ }\href@noop {} {\bibfield  {journal} {\bibinfo
  {journal} {Biophysical journal}\ }\textbf {\bibinfo {volume} {114}},\
  \bibinfo {pages} {1667} (\bibinfo {year} {2018})}\BibitemShut {NoStop}%
\bibitem [{\citenamefont {Fodor}\ \emph {et~al.}(2018)\citenamefont {Fodor},
  \citenamefont {Mehandia}, \citenamefont {Comelles}, \citenamefont
  {Thiagarajan}, \citenamefont {Gov}, \citenamefont {Visco}, \citenamefont {van
  Wijland},\ and\ \citenamefont {Riveline}}]{fodor2018spatial}%
  \BibitemOpen
  \bibfield  {author} {\bibinfo {author} {\bibfnamefont {{\'E}.}~\bibnamefont
  {Fodor}}, \bibinfo {author} {\bibfnamefont {V.}~\bibnamefont {Mehandia}},
  \bibinfo {author} {\bibfnamefont {J.}~\bibnamefont {Comelles}}, \bibinfo
  {author} {\bibfnamefont {R.}~\bibnamefont {Thiagarajan}}, \bibinfo {author}
  {\bibfnamefont {N.~S.}\ \bibnamefont {Gov}}, \bibinfo {author} {\bibfnamefont
  {P.}~\bibnamefont {Visco}}, \bibinfo {author} {\bibfnamefont
  {F.}~\bibnamefont {van Wijland}},\ and\ \bibinfo {author} {\bibfnamefont
  {D.}~\bibnamefont {Riveline}},\ }\href@noop {} {\bibfield  {journal}
  {\bibinfo  {journal} {Biophysical journal}\ }\textbf {\bibinfo {volume}
  {114}},\ \bibinfo {pages} {939} (\bibinfo {year} {2018})}\BibitemShut
  {NoStop}%
\bibitem [{\citenamefont {Tailleur}\ and\ \citenamefont
  {Cates}(2009)}]{tailleur2009sedimentation}%
  \BibitemOpen
  \bibfield  {author} {\bibinfo {author} {\bibfnamefont {J.}~\bibnamefont
  {Tailleur}}\ and\ \bibinfo {author} {\bibfnamefont {M.}~\bibnamefont
  {Cates}},\ }\href@noop {} {\bibfield  {journal} {\bibinfo  {journal} {EPL
  (Europhysics Letters)}\ }\textbf {\bibinfo {volume} {86}},\ \bibinfo {pages}
  {60002} (\bibinfo {year} {2009})}\BibitemShut {NoStop}%
\bibitem [{\citenamefont {Sevilla}\ \emph {et~al.}(2019)\citenamefont
  {Sevilla}, \citenamefont {Arzola},\ and\ \citenamefont
  {Cital}}]{sevilla2019stationary}%
  \BibitemOpen
  \bibfield  {author} {\bibinfo {author} {\bibfnamefont {F.~J.}\ \bibnamefont
  {Sevilla}}, \bibinfo {author} {\bibfnamefont {A.~V.}\ \bibnamefont
  {Arzola}},\ and\ \bibinfo {author} {\bibfnamefont {E.~P.}\ \bibnamefont
  {Cital}},\ }\href@noop {} {\bibfield  {journal} {\bibinfo  {journal}
  {Physical Review E}\ }\textbf {\bibinfo {volume} {99}},\ \bibinfo {pages}
  {012145} (\bibinfo {year} {2019})}\BibitemShut {NoStop}%
\bibitem [{\citenamefont {Kramers}(1940)}]{Kramer1}%
  \BibitemOpen
  \bibfield  {author} {\bibinfo {author} {\bibfnamefont {H.~A.}\ \bibnamefont
  {Kramers}},\ }\href@noop {} {\bibfield  {journal} {\bibinfo  {journal}
  {Physics}\ }\textbf {\bibinfo {volume} {7}},\ \bibinfo {pages} {204}
  (\bibinfo {year} {1940})}\BibitemShut {NoStop}%
\bibitem [{\citenamefont {H{\"{a}}nggi}\ \emph {et~al.}(1990)\citenamefont
  {H{\"{a}}nggi}, \citenamefont {Talkner},\ and\ \citenamefont
  {Borkovec}}]{Kramer2}%
  \BibitemOpen
  \bibfield  {author} {\bibinfo {author} {\bibfnamefont {P.}~\bibnamefont
  {H{\"{a}}nggi}}, \bibinfo {author} {\bibfnamefont {P.}~\bibnamefont
  {Talkner}},\ and\ \bibinfo {author} {\bibfnamefont {M.}~\bibnamefont
  {Borkovec}},\ }\href@noop {} {\bibfield  {journal} {\bibinfo  {journal} {Rev.
  Mod. Phys.}\ }\textbf {\bibinfo {volume} {62}},\ \bibinfo {pages} {777}
  (\bibinfo {year} {1990})}\BibitemShut {NoStop}%
\bibitem [{\citenamefont {Mel'nikov}(1991)}]{Melnikov1991}%
  \BibitemOpen
  \bibfield  {author} {\bibinfo {author} {\bibfnamefont {V.~I.}\ \bibnamefont
  {Mel'nikov}},\ }\href@noop {} {\bibfield  {journal} {\bibinfo  {journal}
  {Physics Reports}\ }\textbf {\bibinfo {volume} {209}},\ \bibinfo {pages} {1}
  (\bibinfo {year} {1991})}\BibitemShut {NoStop}%
\bibitem [{\citenamefont {Bel}\ and\ \citenamefont
  {Ashkenazy}(2013)}]{Bel2013}%
  \BibitemOpen
  \bibfield  {author} {\bibinfo {author} {\bibfnamefont {G.}~\bibnamefont
  {Bel}}\ and\ \bibinfo {author} {\bibfnamefont {Y.}~\bibnamefont
  {Ashkenazy}},\ }\href@noop {} {\bibfield  {journal} {\bibinfo  {journal} {New
  J. Phys.}\ }\textbf {\bibinfo {volume} {15}},\ \bibinfo {pages} {053024}
  (\bibinfo {year} {2013})}\BibitemShut {NoStop}%
\bibitem [{\citenamefont {Milshtein}\ and\ \citenamefont
  {Tret'yakov}(1994)}]{Milshtein1994}%
  \BibitemOpen
  \bibfield  {author} {\bibinfo {author} {\bibfnamefont {G.~N.}\ \bibnamefont
  {Milshtein}}\ and\ \bibinfo {author} {\bibfnamefont {M.~V.}\ \bibnamefont
  {Tret'yakov}},\ }\href {https://doi.org/10.1007/BF02179457} {\bibfield
  {journal} {\bibinfo  {journal} {Journal of Statistical Physics}\ }\textbf
  {\bibinfo {volume} {77}},\ \bibinfo {pages} {691} (\bibinfo {year}
  {1994})}\BibitemShut {NoStop}%
\bibitem [{\citenamefont {Marchetti}\ \emph
  {et~al.}(2013{\natexlab{b}})\citenamefont {Marchetti}, \citenamefont
  {Joanny}, \citenamefont {Ramaswamy}, \citenamefont {Liverpool}, \citenamefont
  {Prost}, \citenamefont {Rao},\ and\ \citenamefont {Simha}}]{Marchetti2013}%
  \BibitemOpen
  \bibfield  {author} {\bibinfo {author} {\bibfnamefont {M.~C.}\ \bibnamefont
  {Marchetti}}, \bibinfo {author} {\bibfnamefont {J.~F.}\ \bibnamefont
  {Joanny}}, \bibinfo {author} {\bibfnamefont {S.}~\bibnamefont {Ramaswamy}},
  \bibinfo {author} {\bibfnamefont {T.~B.}\ \bibnamefont {Liverpool}}, \bibinfo
  {author} {\bibfnamefont {J.}~\bibnamefont {Prost}}, \bibinfo {author}
  {\bibfnamefont {M.}~\bibnamefont {Rao}},\ and\ \bibinfo {author}
  {\bibfnamefont {R.~A.}\ \bibnamefont {Simha}},\ }\href
  {https://doi.org/10.1103/RevModPhys.85.1143} {\bibfield  {journal} {\bibinfo
  {journal} {Rev. Mod. Phys.}\ }\textbf {\bibinfo {volume} {85}},\ \bibinfo
  {pages} {1143} (\bibinfo {year} {2013}{\natexlab{b}})}\BibitemShut {NoStop}%
\bibitem [{\citenamefont {Sonn-Segev}\ \emph {et~al.}(2017)\citenamefont
  {Sonn-Segev}, \citenamefont {Bernheim-Groswasser},\ and\ \citenamefont
  {Roichman}}]{Son2017}%
  \BibitemOpen
  \bibfield  {author} {\bibinfo {author} {\bibfnamefont {A.}~\bibnamefont
  {Sonn-Segev}}, \bibinfo {author} {\bibfnamefont {A.}~\bibnamefont
  {Bernheim-Groswasser}},\ and\ \bibinfo {author} {\bibfnamefont
  {Y.}~\bibnamefont {Roichman}},\ }\href {https://doi.org/10.1039/C7SM01391D}
  {\bibfield  {journal} {\bibinfo  {journal} {Soft Matter}\ }\textbf {\bibinfo
  {volume} {13}},\ \bibinfo {pages} {7352} (\bibinfo {year}
  {2017})}\BibitemShut {NoStop}%
\bibitem [{\citenamefont {Elgeti}\ \emph {et~al.}(2015)\citenamefont {Elgeti},
  \citenamefont {Winkler},\ and\ \citenamefont {Gompper}}]{Elgeti2015}%
  \BibitemOpen
  \bibfield  {author} {\bibinfo {author} {\bibfnamefont {J.}~\bibnamefont
  {Elgeti}}, \bibinfo {author} {\bibfnamefont {R.~G.}\ \bibnamefont
  {Winkler}},\ and\ \bibinfo {author} {\bibfnamefont {G.}~\bibnamefont
  {Gompper}},\ }\href {https://doi.org/10.1088/0034-4885/78/5/056601}
  {\bibfield  {journal} {\bibinfo  {journal} {Reports on Progress in Physics}\
  }\textbf {\bibinfo {volume} {78}},\ \bibinfo {pages} {056601} (\bibinfo
  {year} {2015})}\BibitemShut {NoStop}%
\bibitem [{\citenamefont {Palacci}\ \emph {et~al.}(2010)\citenamefont
  {Palacci}, \citenamefont {Cottin-Bizonne}, \citenamefont {Ybert},\ and\
  \citenamefont {Bocquet}}]{Palacci2010}%
  \BibitemOpen
  \bibfield  {author} {\bibinfo {author} {\bibfnamefont {J.}~\bibnamefont
  {Palacci}}, \bibinfo {author} {\bibfnamefont {C.}~\bibnamefont
  {Cottin-Bizonne}}, \bibinfo {author} {\bibfnamefont {C.}~\bibnamefont
  {Ybert}},\ and\ \bibinfo {author} {\bibfnamefont {L.}~\bibnamefont
  {Bocquet}},\ }\href@noop {} {\bibfield  {journal} {\bibinfo  {journal}
  {Physical Review Letters}\ }\textbf {\bibinfo {volume} {105}},\ \bibinfo
  {pages} {088304} (\bibinfo {year} {2010})}\BibitemShut {NoStop}%
\bibitem [{\citenamefont {Ginot}\ \emph {et~al.}(2015)\citenamefont {Ginot},
  \citenamefont {Theurkauff}, \citenamefont {Levis}, \citenamefont {Ybert},
  \citenamefont {Bocquet}, \citenamefont {Berthier},\ and\ \citenamefont
  {Cottin-Bizonne}}]{Ginot2015}%
  \BibitemOpen
  \bibfield  {author} {\bibinfo {author} {\bibfnamefont {F.}~\bibnamefont
  {Ginot}}, \bibinfo {author} {\bibfnamefont {I.}~\bibnamefont {Theurkauff}},
  \bibinfo {author} {\bibfnamefont {D.}~\bibnamefont {Levis}}, \bibinfo
  {author} {\bibfnamefont {C.}~\bibnamefont {Ybert}}, \bibinfo {author}
  {\bibfnamefont {L.}~\bibnamefont {Bocquet}}, \bibinfo {author} {\bibfnamefont
  {L.}~\bibnamefont {Berthier}},\ and\ \bibinfo {author} {\bibfnamefont
  {C.}~\bibnamefont {Cottin-Bizonne}},\ }\href@noop {} {\bibfield  {journal}
  {\bibinfo  {journal} {Physical Review X}\ }\textbf {\bibinfo {volume} {5}},\
  \bibinfo {pages} {011004} (\bibinfo {year} {2015})}\BibitemShut {NoStop}%
\bibitem [{\citenamefont {Levis}\ and\ \citenamefont
  {Berthier}(2015)}]{Levis2015}%
  \BibitemOpen
  \bibfield  {author} {\bibinfo {author} {\bibfnamefont {D.}~\bibnamefont
  {Levis}}\ and\ \bibinfo {author} {\bibfnamefont {L.}~\bibnamefont
  {Berthier}},\ }\href@noop {} {\bibfield  {journal} {\bibinfo  {journal} {EPL
  (Europhysics Letters)}\ }\textbf {\bibinfo {volume} {111}},\ \bibinfo {pages}
  {60006} (\bibinfo {year} {2015})}\BibitemShut {NoStop}%
\bibitem [{\citenamefont {Han}\ \emph {et~al.}(2017)\citenamefont {Han},
  \citenamefont {Yan}, \citenamefont {Granick},\ and\ \citenamefont
  {Luijten}}]{Han2017}%
  \BibitemOpen
  \bibfield  {author} {\bibinfo {author} {\bibfnamefont {M.}~\bibnamefont
  {Han}}, \bibinfo {author} {\bibfnamefont {J.}~\bibnamefont {Yan}}, \bibinfo
  {author} {\bibfnamefont {S.}~\bibnamefont {Granick}},\ and\ \bibinfo {author}
  {\bibfnamefont {E.}~\bibnamefont {Luijten}},\ }\href
  {https://doi.org/10.1073/pnas.1706702114} {\bibfield  {journal} {\bibinfo
  {journal} {Proceedings of the National Academy of Sciences}\ }\textbf
  {\bibinfo {volume} {114}},\ \bibinfo {pages} {7513} (\bibinfo {year}
  {2017})}\BibitemShut {NoStop}%
\bibitem [{\citenamefont {Wittmann}\ \emph {et~al.}(2018)\citenamefont
  {Wittmann}, \citenamefont {Brader}, \citenamefont {Sharma},\ and\
  \citenamefont {Marconi}}]{Wittmann2018}%
  \BibitemOpen
  \bibfield  {author} {\bibinfo {author} {\bibfnamefont {R.}~\bibnamefont
  {Wittmann}}, \bibinfo {author} {\bibfnamefont {J.~M.}\ \bibnamefont
  {Brader}}, \bibinfo {author} {\bibfnamefont {A.}~\bibnamefont {Sharma}},\
  and\ \bibinfo {author} {\bibfnamefont {U.~M.~B.}\ \bibnamefont {Marconi}},\
  }\href {https://doi.org/10.1103/PhysRevE.97.012601} {\bibfield  {journal}
  {\bibinfo  {journal} {Phys. Rev. E}\ }\textbf {\bibinfo {volume} {97}},\
  \bibinfo {pages} {012601} (\bibinfo {year} {2018})}\BibitemShut {NoStop}%
\bibitem [{\citenamefont {Bechinger}\ \emph {et~al.}(2016)\citenamefont
  {Bechinger}, \citenamefont {Di~Leonardo}, \citenamefont {L\"owen},
  \citenamefont {Reichhardt}, \citenamefont {Volpe},\ and\ \citenamefont
  {Volpe}}]{Bechinger2016}%
  \BibitemOpen
  \bibfield  {author} {\bibinfo {author} {\bibfnamefont {C.}~\bibnamefont
  {Bechinger}}, \bibinfo {author} {\bibfnamefont {R.}~\bibnamefont
  {Di~Leonardo}}, \bibinfo {author} {\bibfnamefont {H.}~\bibnamefont
  {L\"owen}}, \bibinfo {author} {\bibfnamefont {C.}~\bibnamefont {Reichhardt}},
  \bibinfo {author} {\bibfnamefont {G.}~\bibnamefont {Volpe}},\ and\ \bibinfo
  {author} {\bibfnamefont {G.}~\bibnamefont {Volpe}},\ }\href
  {https://doi.org/10.1103/RevModPhys.88.045006} {\bibfield  {journal}
  {\bibinfo  {journal} {Rev. Mod. Phys.}\ }\textbf {\bibinfo {volume} {88}},\
  \bibinfo {pages} {045006} (\bibinfo {year} {2016})}\BibitemShut {NoStop}%
\bibitem [{\citenamefont {Ni}\ \emph {et~al.}(2013)\citenamefont {Ni},
  \citenamefont {Stuart},\ and\ \citenamefont {Dijkstra}}]{Ni2013}%
  \BibitemOpen
  \bibfield  {author} {\bibinfo {author} {\bibfnamefont {R.}~\bibnamefont
  {Ni}}, \bibinfo {author} {\bibfnamefont {M.~A.~C.}\ \bibnamefont {Stuart}},\
  and\ \bibinfo {author} {\bibfnamefont {M.}~\bibnamefont {Dijkstra}},\
  }\href@noop {} {\bibfield  {journal} {\bibinfo  {journal} {Nature
  communications}\ }\textbf {\bibinfo {volume} {4}},\ \bibinfo {pages} {2704}
  (\bibinfo {year} {2013})}\BibitemShut {NoStop}%
\bibitem [{\citenamefont {Pollak}(1990)}]{Pollak1990}%
  \BibitemOpen
  \bibfield  {author} {\bibinfo {author} {\bibfnamefont {E.}~\bibnamefont
  {Pollak}},\ }\href@noop {} {\bibfield  {journal} {\bibinfo  {journal} {J.
  Chem. Phys.}\ }\textbf {\bibinfo {volume} {93}},\ \bibinfo {pages} {1116}
  (\bibinfo {year} {1990})}\BibitemShut {NoStop}%
\end{thebibliography}%

\clearpage
\appendix
\section{The active force implementations}
For an integer number of sources,
$$ \varphi_N\left(\tilde{t}\right)\equiv\frac{1}{\sqrt{N}}\displaystyle\sum\limits_{r=1}^N \varphi_{1,r}\left(\tilde{t}\right) $$
where the time dependence of each source is given by
\begin{align}\label{eq:singmot}
\varphi_{1,r}\left(\tilde{t}\right) = \begin{cases}
    0 & \exists i:\,t\in\mathcal{T}_{off,i,r}\\
    a_{i,r} & \exists i:\,t\in\mathcal{T}_{on,i,r},
    \end{cases}
\end{align}
where $a_{i,r}=\pm1$ with equal probability ($a_{i,r}$ is drawn independently for each interval of "on" time), and the duration of each "on"/"off" interval is drawn from an exponential distribution according to
\begin{align}\label{eq:ootpdf}
    p\left(\mathcal{T}_{on,i,r}\right)&=\tau_{on}^{-1}exp\left(-\mathcal{T}_{on,i,r}/\tau_{on}\right); \\
    p\left(\mathcal{T}_{off,i,r}\right)&=\tau_{off}^{-1}exp\left(-\mathcal{T}_{off,i,r}/\tau_{off}\right).
\end{align}
It is important to note that the different sources are independent in the sense that each source has different "on" and "off" periods and an independent direction of the force ($a_{i,r}$ for different sources ($r$ values) are independent). All sources share the same statistics regarding the duration of the intervals and the relative contribution to the total active force. It is worth noting that the fixed initial time, $t=0$, the duration of the intervals for each source, and their sequential order completely define the time series describing the active force values.

For the Gaussian implementation of the active force, which corresponds to the limit of a large number of sources, $N\to\infty$, $\varphi_G\left(\tilde{t}\right)$ is a Gaussian colored noise with $\langle\varphi_G\left(\tilde{t}\right)\varphi_G\left(\tilde{t}'\right)\rangle=p_{on}exp\left(-|\tilde{t}-\tilde{t}'|/(\tau_{on}/t_d)\right)$. The specific implementation is realized as an Ornstein-Uhlenbeck process according to:
\begin{equation}\label{eq:fadyn}
\frac{d\varphi_{G}(\tilde{t})}{d\tilde{t}}=-\frac{1}{\tau_{on}/t_d}\varphi_G(\tilde{t})+\eta(\tilde{t}),
\end{equation}
where $\eta(\tilde{t})$ is a Gaussian white noise whose characteristics are given by:
$$    \langle \eta(\tilde{t})\rangle=0,$$
and
$$\langle\eta(\tilde{t})\eta(\tilde{t}')\rangle=\frac{2p_{on}}{\tau_{on}/t_d}\delta(\tilde{t}-\tilde{t}').$$

\renewcommand{\thefigure}{{B.\arabic{figure}}}
\section{Position and velocity probability density functions for trapped particles under the influence of the Gaussian implementation of the active force}\label{app:A}
\setcounter{figure}{0}

\begin{figure*}[b]
\includegraphics[width=\linewidth,trim= 3.5cm 0cm 5cm 1cm,clip]{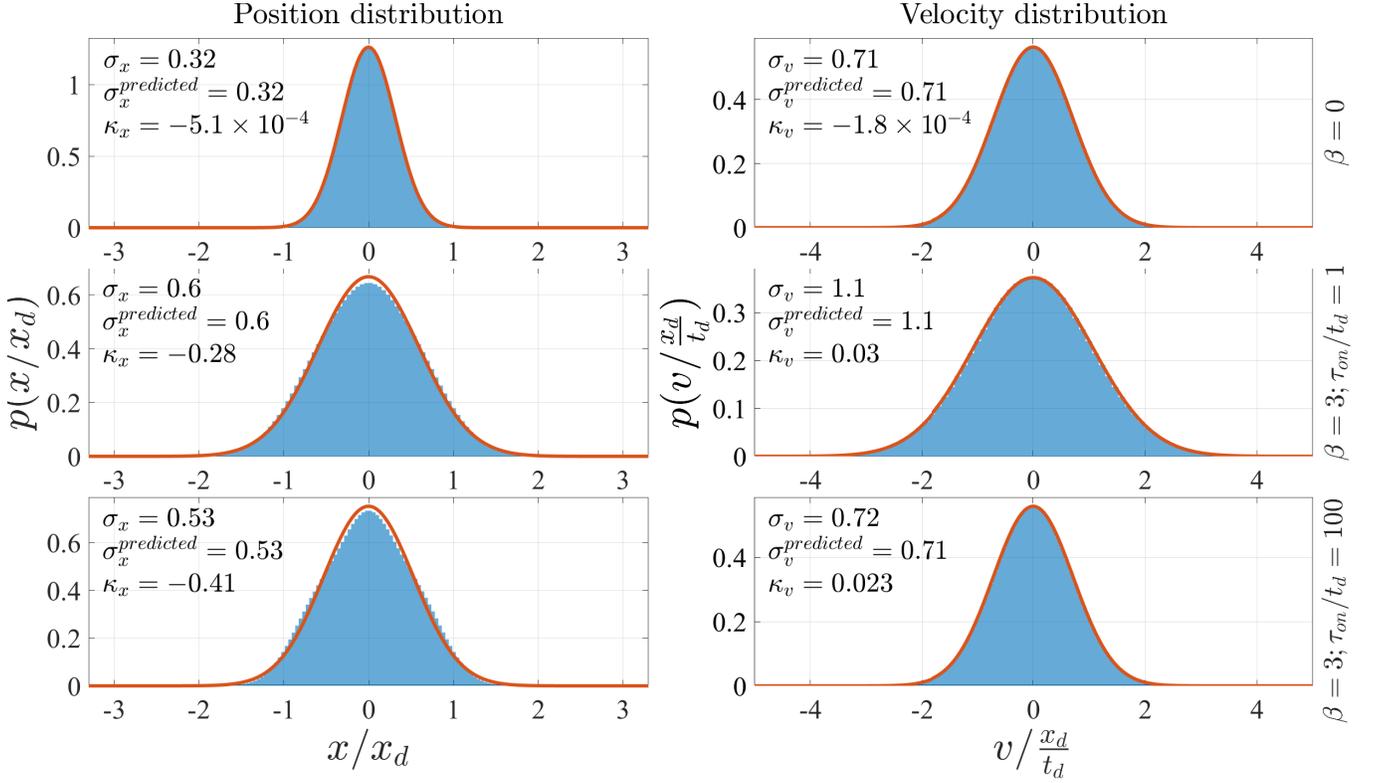}
\caption{\label{fig:nehist1} Simulated probability density functions of the position and the velocity of particles within an infinite harmonic trap ($\alpha=5$). The red lines represent the corresponding Gaussian distributions expected for a thermal particle within a harmonic trap with the same parameters and a temperature given by the effective temperature $T_x$ or $T_v$, respectively. Due to the relatively weak active force, all the distributions are uni-modal, and their width is in excellent agreement with the predicted width. The active force is characterized by $p_{on}=0.5$.}
\end{figure*}
\begin{figure*}[b]
\includegraphics[width=\linewidth,trim=3.5cm 0cm 5cm 0cm,clip]{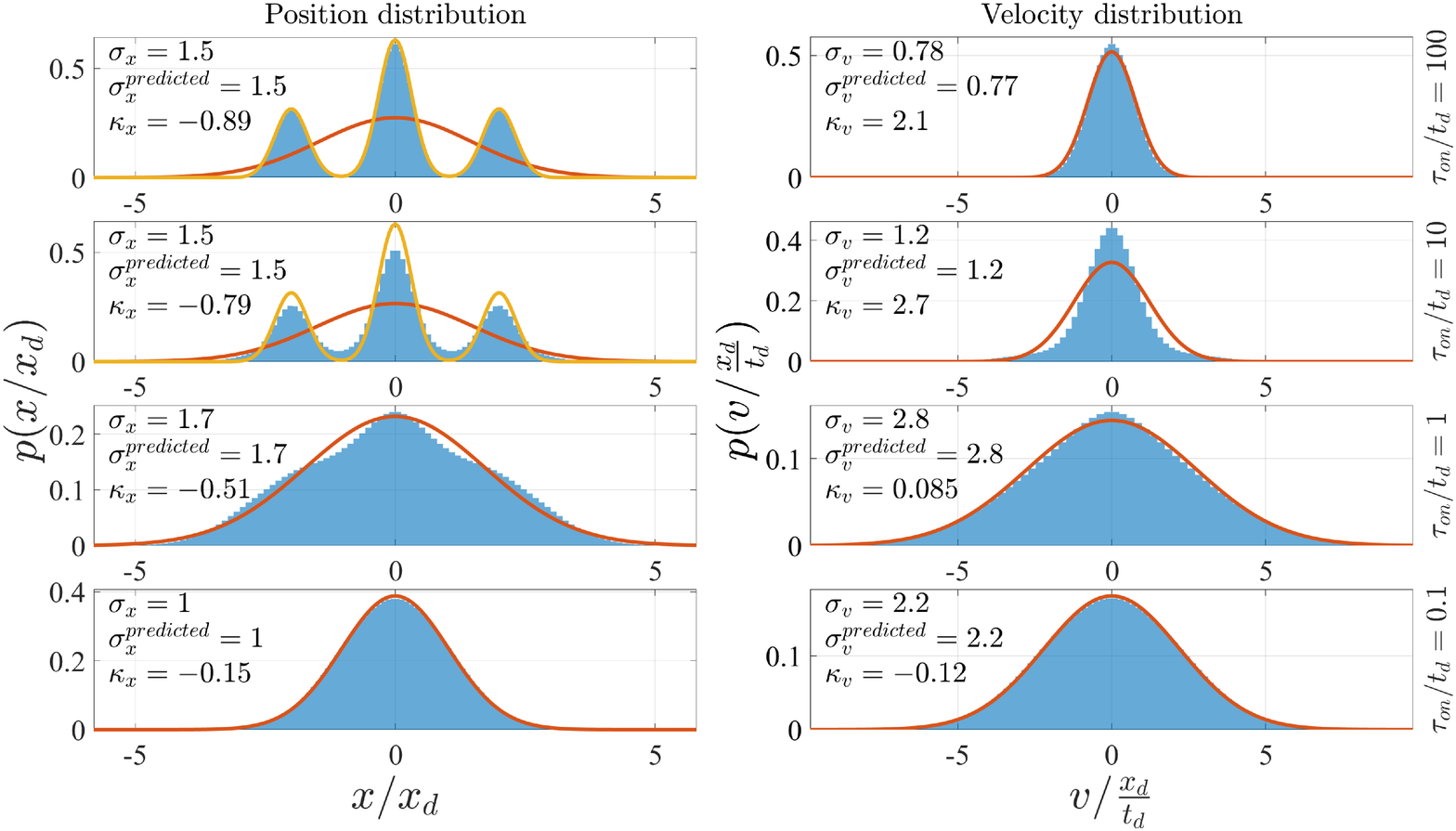}
\caption{\label{fig:nehist2} Similar to Fig. \ref{fig:nehist1} but for a stronger active force, $\beta=10$ ($\alpha=5$ and $p_{on}=0.5$). For short correlation times of the active force, bottom panels, the same behavior found for the weaker force is seen (i.e., Gaussian distributions). For longer correlation times, the position probability density function becomes tri-modal due to the time spent near the points where the active force balances the force due to the harmonic potential. At these points, the velocity is small, and therefore, the corresponding velocity PDFs are narrower than Gaussian.  Note that even for the cases with non-Gaussian PDFs, the predicted second moments are still in agreement with the simulated ones. The yellow lines fitting the tri-modal position PDFs are detailed in the text.}
\end{figure*}

To better characterize the dynamics of the trapped active particles, we derived their position and velocity probability density functions (PDFs).
The histograms of the velocity and position of the trapped particles were derived from simulated trajectories. The convergence was verified by comparing the histograms derived from different trajectories with the requirement that $(1/2)\intop_{-\infty}^\infty|\psi_{i}(q)-\psi_{j}(q)| \mathrm{d}q<0.01$, where $\psi_{i}(q)$ is the PDF of the variable $q$ ($x$ or $v$) derived from the $i$'th simulated trajectory. The simulations were done with a time step of $dt=min(\tau_{on} / 30, 0.01)$, and we found that trajectories of $10^{10}$ steps (equivalent to a total duration of $1/3$\textendash$1 \times 10^8 t_d$) satisfied the convergence condition.\\
In order to quantify the deviation from a Gaussian distribution, we show for each probability density function (PDF), its standard deviation (STD), $\sigma$, the calculated STD, $\sigma^{predicted}$ (based on eqs. \eqref{eq:T_eff_x} or \eqref{eq:T_eff_v}), and the excess kurtosis, $\kappa$.  
As expected, for small values of the active force amplitude, the PDFs are close to the thermal Gaussian PDF (not shown). For all the parameter values, we found an excellent agreement between the calculated and simulated second moments of the position and the velocity.

\begin{figure*}[b]
\includegraphics[width=\linewidth]{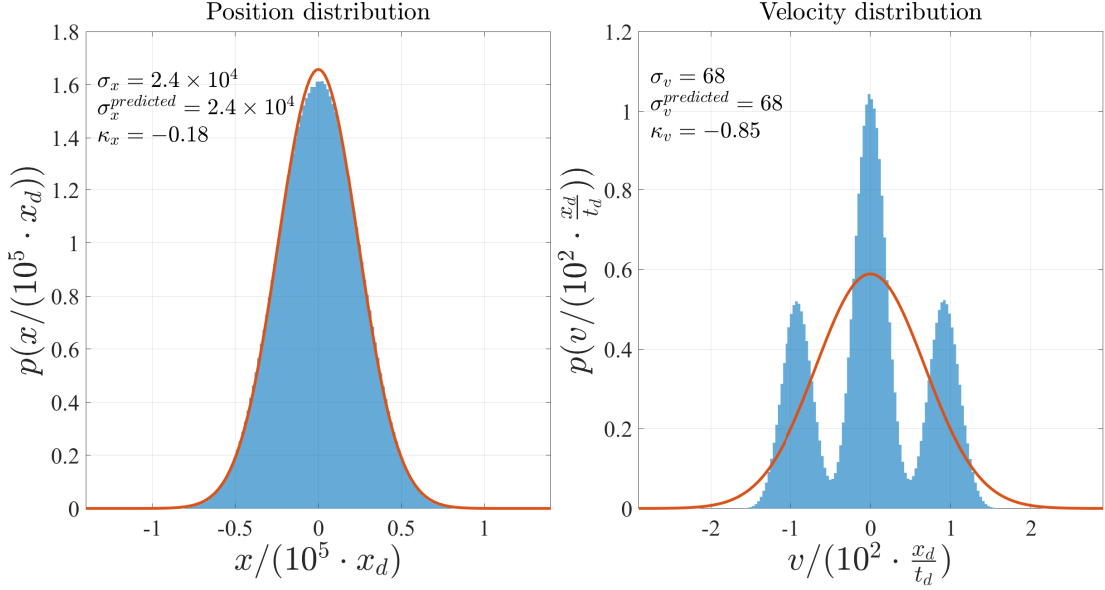}
\caption{\label{fig:nehist3} Similar to Figures \ref{fig:nehist1} and \ref{fig:nehist2} but with a much shallower trap ($\alpha=0.0008$). The correlation time is $\tau_{on}=100t_d$, the force amplitude is $\beta=100$, and $p_{on}=0.5$. In this case, the position PDF is close to Gaussian, but the velocity PDF is tri-modal. The tri-modality stems from the periods of ballistic motion due to the active force (the trap has little influence near its bottom due to the weak confinement).}
\end{figure*}

In Figs. \ref{fig:nehist1}\textendash\ref{fig:nehist3}, we present histograms of the position and velocity for different values of a single source active force amplitude and auto-correlation time.
In Fig. \ref{fig:nehist1}, we present the PDFs for the thermal case (upper row, $\beta=0$) and for an active force with a small amplitude, $\beta=3$, and two different auto-correlation times of the active force, $\tau_{on}=1$ (middle row) and $\tau_{on}=100$ (lowest row). The thermal case is presented just to enable a comparison. All the panels correspond to harmonic potential with $\alpha=5$, and the single source active force is characterized by $p_{on}=0.5$. For the small amplitude active force, one can see that the PDFs of the position are uni-modal for both correlation times, but the PDF is not Gaussian; the negative excess kurtosis implies that extreme values of the position are less likely than in a Gaussian PDF. The velocity PDFs are very close to Gaussian.

In Fig. \ref{fig:nehist2}, we present the PDFs of the position and velocity but for a larger force amplitude, $\beta=10$. Due to the larger active force, there are stationary points away from the bottom of the potential well, and for long auto-correlation times of the active force, the position PDF becomes tri-modal. The stationary points are determined by $\alpha |\tilde{x}|=\beta$. The width of each peak is determined by the thermal fluctuations ($T$) around the stationary points, and the relative height of the side bands is determined by $p_{on}$ (the fraction of time during which the active force is 'on"). For the specific parameters used in Fig. \ref{fig:nehist2}, the yellow line shows the following function,
\begin{widetext}
$$p(x)=\frac{1}{\sqrt{2\pi \langle x^2\rangle_{thermal}}}\left(\frac{p_{on}}{2}\left[exp\left(- \frac{(x+x_s)^2}{2 \langle x^2\rangle_{thermal}}\right)+exp\left(- \frac{(x-x_s)^2}{2 \langle x^2\rangle_{thermal}}\right)\right]+(1-p_{on})exp\left(- \frac{x^2}{2 \langle x^2\rangle_{thermal}}\right)\right) $$.
\end{widetext}
$x_s=\beta/\alpha x_d=2 x_d$ denotes the absolute value of the stationary points, and $\langle x^2\rangle_{thermal}=k_B T/k=x_d^2/(2\alpha)=0.1 x_d^2$ denotes the variance of position fluctuations under thermal force (the latter equality stems from the parameters corresponding to Fig. \ref{fig:nehist2}).
Despite the tri-modal PDFs, the second moments are still in agreement with the calculated ones since we have not made any assumption regarding the shape of the PDF in deriving the second moments. It is worth noting that around the stationary points, the velocity fluctuations are still around zero; therefore, for these parameters, the velocity PDFs are uni-modal.\\

\begin{figure*}[b]
\includegraphics[width=\linewidth]{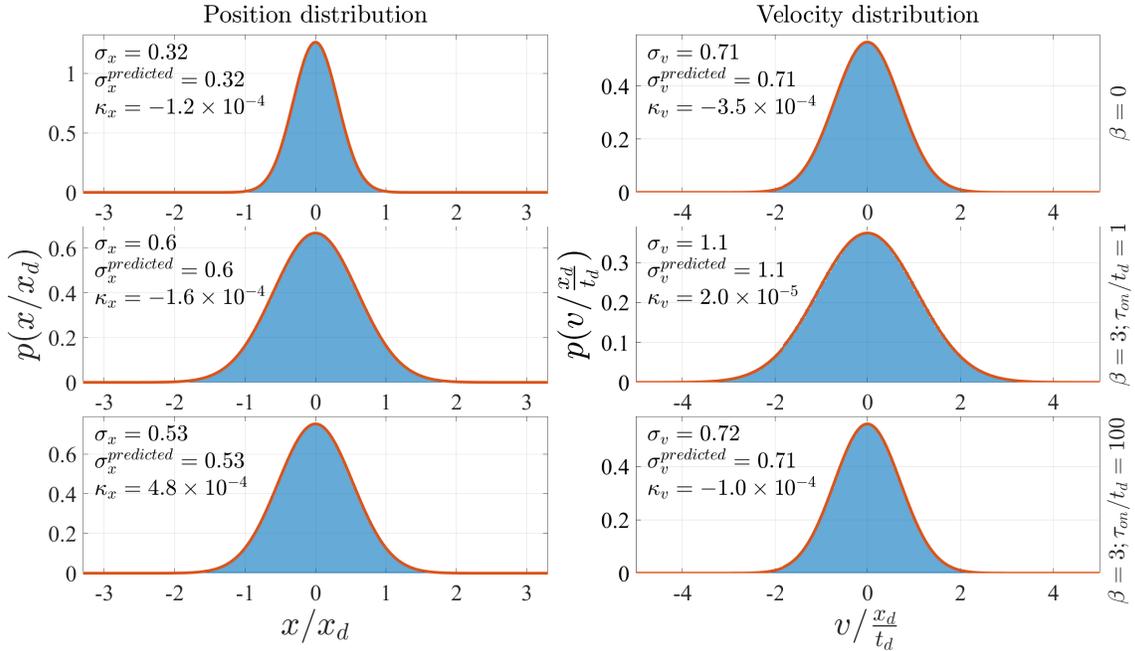}
\caption{\label{fig:nehistG1}
Probability density functions of the position and velocity for relatively weak active forces. The red lines present Gaussian PDFs with zero mean and variance according to eqs. \eqref{eq:T_eff_x} and \eqref{eq:T_eff_v}. The upper row presents the PDFs for the thermal case, $\beta=0$. The middle and bottom rows represent the PDFs for $\beta=3$ and $\tau_{on}=1t_d$ and $100t_d$, respectively. For all panels, $\alpha=5$.
}
\end{figure*}

\begin{figure*}[b]
\includegraphics[width=\linewidth]{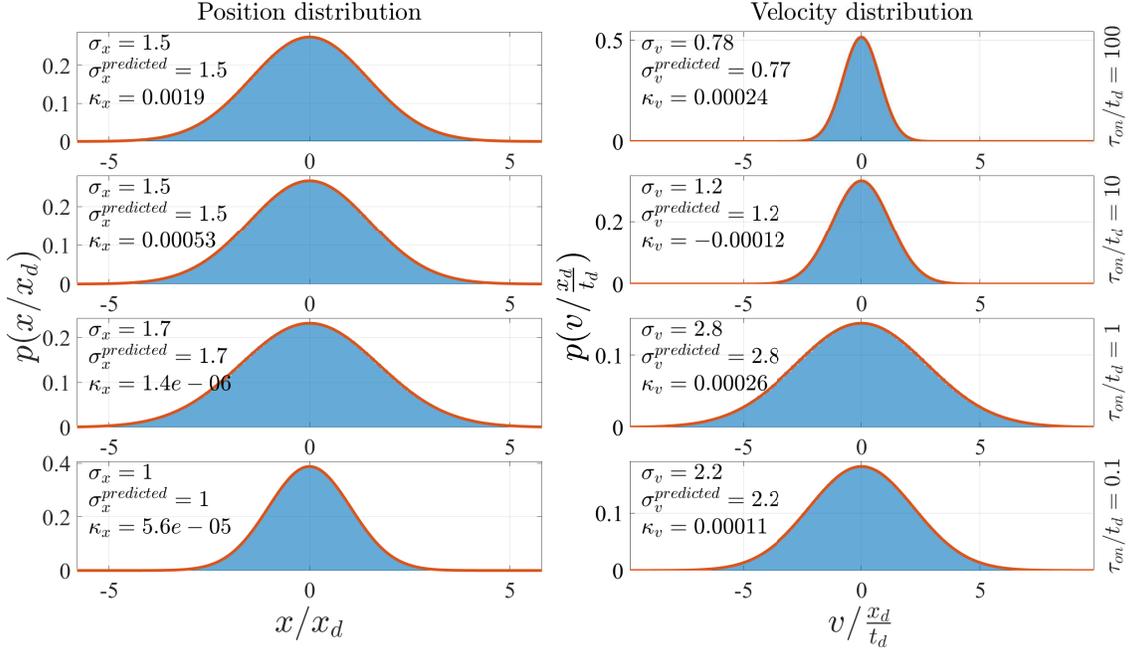}
\caption{\label{fig:nehistG2}
Probability density functions of the position and velocity for a larger amplitude of the active force, $\beta=10$. The red lines present Gaussian PDFs with zero mean and variance according to eqs. \eqref{eq:T_eff_x} and \eqref{eq:T_eff_v}. The different rows correspond to the different auto-correlation times indicated. For all panels, $\alpha=5$.
}
\end{figure*}

\begin{figure*}[b]
\includegraphics[width=\linewidth]{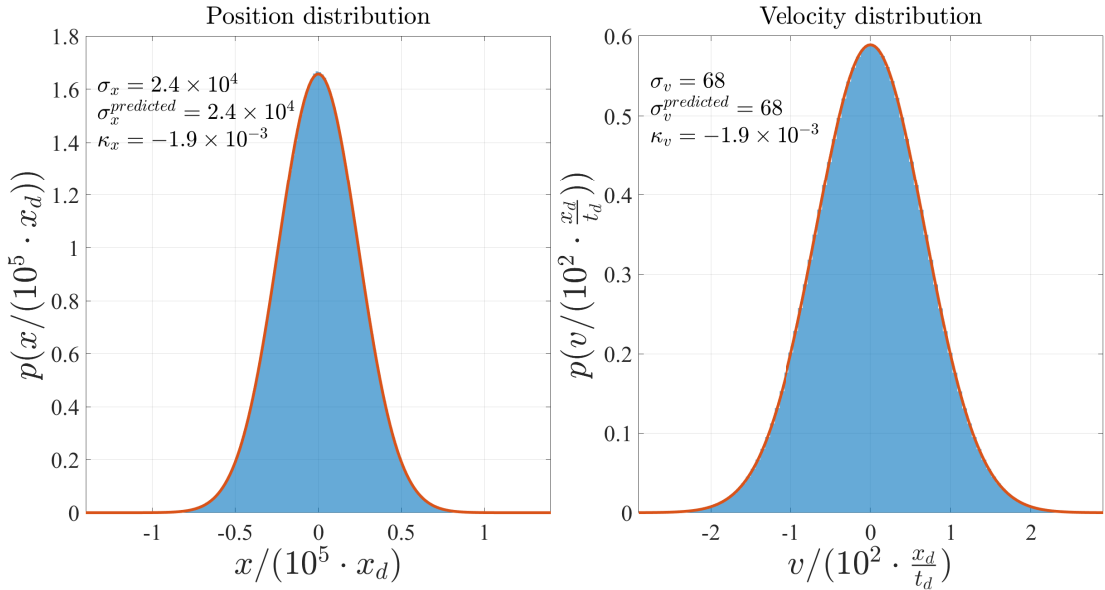}
\caption{\label{fig:nehistG3}
Probability density functions of the position and velocity for a very large amplitude of the Gaussian active force and weak confinement by the potential. The red lines present Gaussian PDFs with zero mean and variance according to eqs. \eqref{eq:T_eff_x} and \eqref{eq:T_eff_v}. As expected, the PDFs are very broad for this setting but are still Gaussian with the corresponding effective temperatures. The parameters are: $\alpha=0.0008$, $\tau_{on}=100$ and $\beta=100$.
}
\end{figure*}

For a weaker confinement, $\alpha=0.0008$, a relatively large forcing amplitude, $\beta=100$, and a long auto-correlation time of the active force, $\tau_{on}=100t_d$, the velocity PDF becomes tri-modal, while the position PDF is uni-modal. The PDFs are depicted in Fig. \ref{fig:nehist3}. The tri-modality of the velocity PDF stems from the terminal velocities under the action of the force and the damping. The weak confinement makes the confining force small compared with the other terms. The position PDF is very broad under these conditions. The calculated second moments are also in agreement with the simulated ones for these parameters.
Figures \ref{fig:nehistG1}\textendash\ref{fig:nehistG3} present the position and velocity PDFs for the Gaussian implementation of the active force. For this implementation, the PDFs are all Gaussian and uni-modal. Unlike the single source implementation, for the Gaussian active force, the amplitude of the force varies and is not fixed. There are no "on" and "off" times. Therefore, there is no tri-modality even for the long auto correlation times. The active force amplitude, dictated by $\beta$, and the auto-correlation time, $\tau_{on}$, as well as the potential shape, dictated by $alpha$, affect the variance of the PDFs. For all values considered, there is an excellent agreement between the calculated second moments, eqs. \eqref{eq:T_eff_x} and \eqref{eq:T_eff_v}, and the simulated ones.
\clearpage

\renewcommand{\thefigure}{{C.\arabic{figure}}}
\section{Position and velocity probability density functions for particles escaping the trap}\label{app:B}
\setcounter{figure}{0}

\begin{figure*}[b]
\includegraphics[width=\linewidth]{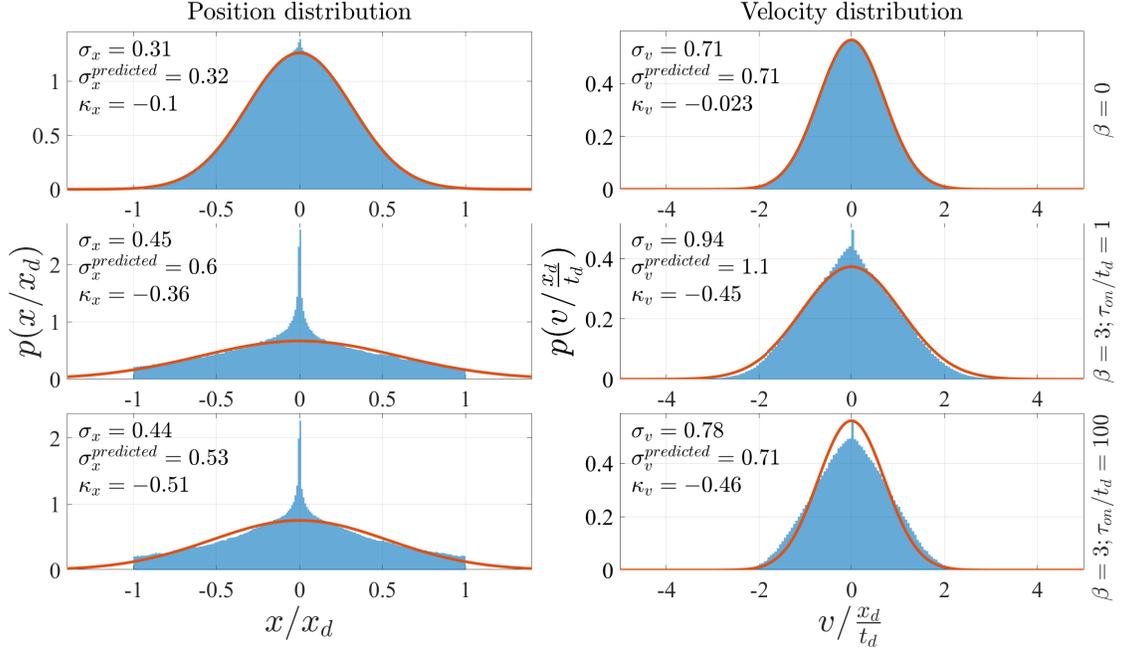}
\caption{\label{fig:ehist1} Position and velocity histograms for the case of escape. The active force was implemented using the single source. The imposed vanishing probability density at the escape points sharply truncates the position PDF. All the distributions show significant deviation from a Gaussian distribution. The second moment of the position is smaller than the second moment calculated for a trapped particle (as expected, due to the zero probability at coordinates beyond the escape points). For the velocity, there is no such effect, and for two of the three parameter sets considered, it appears that faster particles dominate the distribution, resulting in a second moment larger than that of a trapped particle. For all the panels, $\alpha=5$, $p_{on}=0.5$ and $x_{esc}=x_d$. The values of $\beta$ and $\tau_{on}$ are indicated to the right of each row. The upper row presents the results for the thermal case with no active force in order to better illustrate the effects of the active force.}
\end{figure*}

\begin{figure*}[b]
\includegraphics[width=\linewidth]{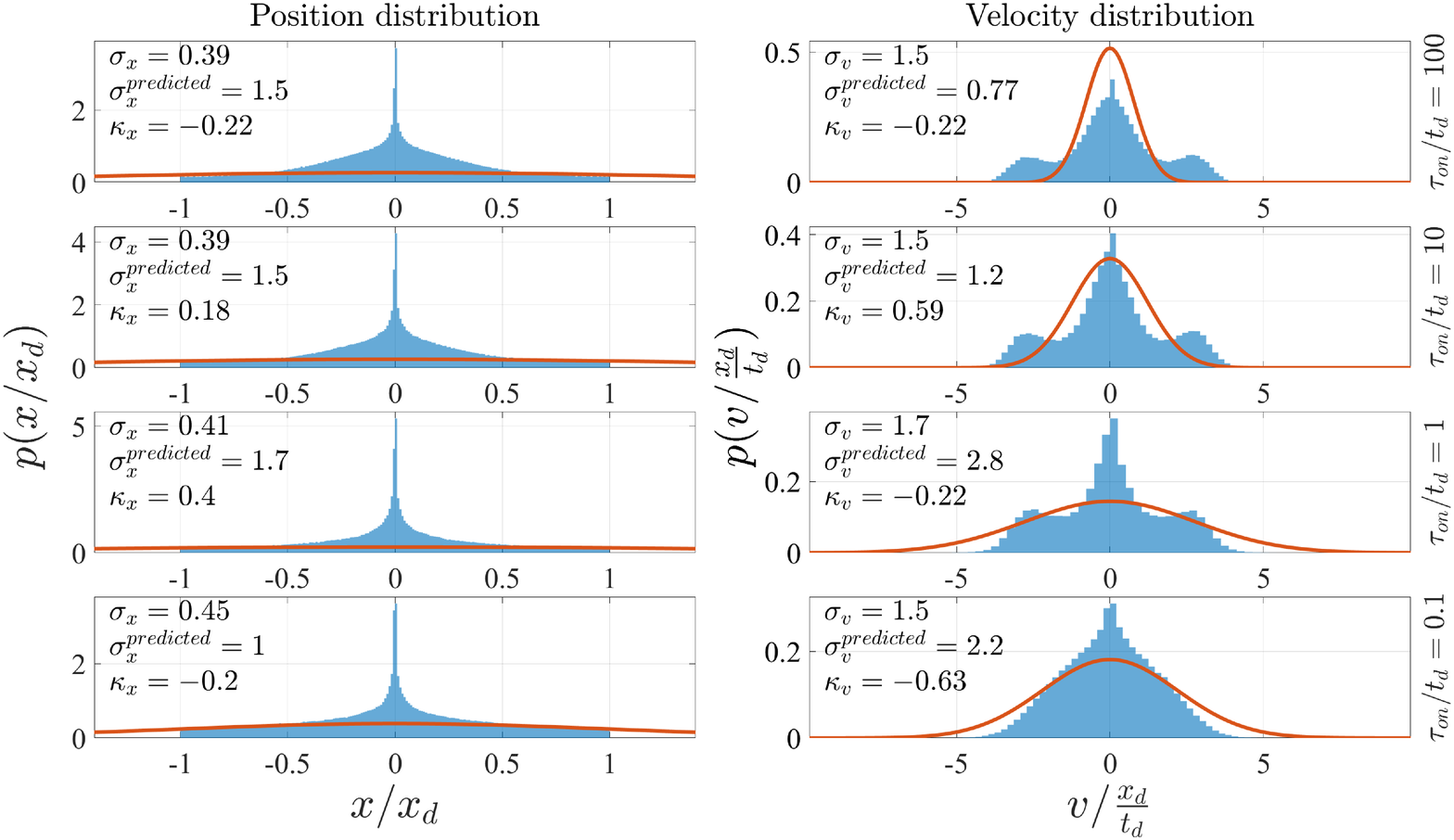}
\caption{\label{fig:ehist2} Similar to Fig. \ref{fig:ehist1} but for $\beta=10$ and different values of $\tau_{on}$. The effect on the position PDFs is similar to that seen in Fig. \ref{fig:ehist1}. These different panels show that for short $\tau_{on}$, the second moment of the velocity is smaller than expected for trapped particles, while for large $\tau_{on}$ it is larger. Moreover, the velocity distribution may become tri-modal for parameters in which trapped particles show uni-modal PDFs. All the PDFs show a clear deviation from a Gaussian PDF. For all panels, $\alpha=5$, $x_{esc}=x_d$ and $p_{on}=0.5$. The values of $\tau_{on}$ are indicated for each row.}
\end{figure*}

\begin{figure*}[b]
\includegraphics[width=\linewidth]{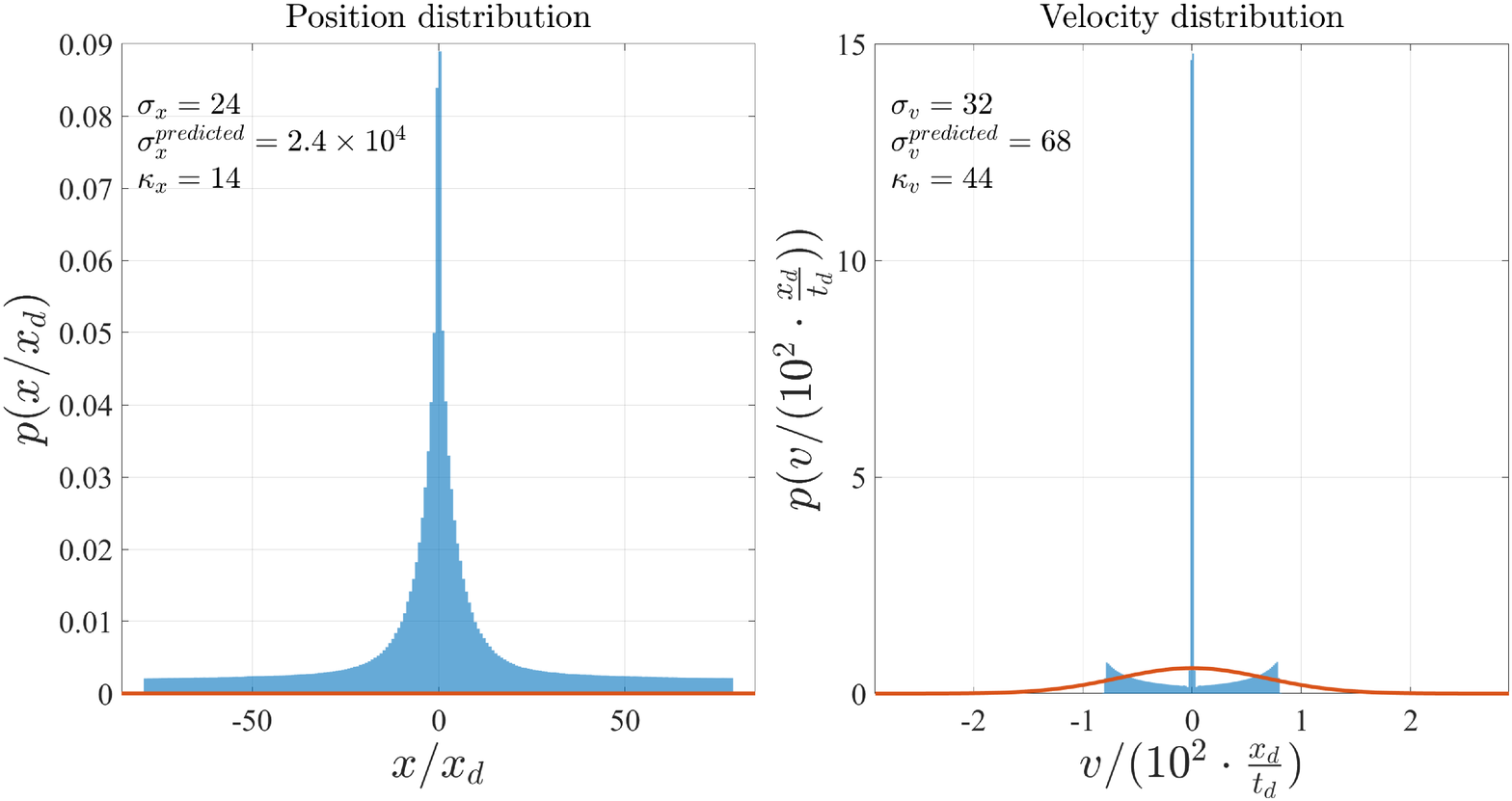}
\caption{\label{fig:ehist3} Position and velocity PDFs for a shallow trap, $\alpha=0.0008$, with escape. Due to the weak confinement, the PDFs are broader. The high peak in the center is likely to be a bias introduced by the initial condition ($x=0$ and $v=0$). The active force parameters are $\beta=100$ and $p_{on}=0.5$. The escape points are $|x_{esc}|=x_d$.}
\end{figure*}

The escape of particles from the well can be considered if one sets an absorbing boundary condition for $|x|=x_{esc}$. Under the settings of our problem, each particle eventually reaches the absorbing boundary and escapes the potential well (see Fig. \ref{fig:etraj} for typical trajectories of a particle that can escape the potential well).
Figures \ref{fig:ehist1}, \ref{fig:ehist2} and \ref{fig:ehist3} present the simulated position and velocity PDFs for particles subjected to a single source active force that can escape the potential well as their position crosses the well boundary $|x|>|x_{esc}$. For particles that escape we expect the PDFs to be different from those of trapped particles. The main difference is that by definition the probability of the particle being outside the well is zero (it is "absorbed" when reaching the boundary).

The PDFs for particles that can escape have to be derived differently because each trajectory terminates once the particle escapes (reaches $|x|=x_{esc}$). To overcome this limitation, we considered multiple trajectories, $4000$ for most values of $\alpha$ and $40000$ for $\alpha=0.0008$. All the trajectories were terminated with the escape of the particle. The time step was identical to the time step used for trapped particles. The PDFs were derived by assigning each trajectory an equal weight regardless of the duration of the trajectory (the weight of each time step was set by $(1/N_T)*(1/T_i)$ where $N_T$ is the number of trajectories that were used and $T_i$ is the duration of the $i$'th trajectory). Therefore, we avoided bias from slowly escaping particles whose trajectories are of longer duration.  
The convergence of the PDFs with escape was verified by considering only a fraction of the total simulated trajectories and verifying that the average difference (quantified by the integral of the absolute value of the difference between the PDFs) between the PDFs derived from two different sets of $N$ trajectories each, is proportional to $1/\sqrt{N}$. The average was taken over $1000$ different sets randomly sampled from the total number of simulated trajectories.
The number of trajectories that we used implies that we expect $\left\langle (1/2)\intop_{-\infty}^\infty|\psi_{i}(q)-\psi_{j}(q)| \mathrm{d}q\right\rangle<0.02$ for PDFs derived from two sets of $N$ trajectories (here, $i,j$ are not trajectory indices but rather set indices, and the total number of trajectories used is $4000$ or $40000$ as mentioned above).

Figure \ref{fig:ehist1} presents the position and velocity PDFs for the same parameters as Fig. \ref{fig:nehist1} but for the case in which the particles escape the potential well. For the thermal case, the main deviation is seen in the position's PDF where the escape of particles results in a deviation from the Gaussian distribution (the excess kurtosis is not zero) and, in particular, in a peak around the bottom of the potential well.
In addition, we found, as expected, that the second moment of the position is smaller than the second moment predicted for trapped particles. The fact that there is a zero probability for the particle to be in the regions $|x|>x_{esc}$ implies that there is no contribution of the tails of the no-escape PDF to the second moment in the case of escape.
The second moment of the velocity can be smaller or larger than the second moment found for the trapped particles and is determined by the force and the potential characteristics ($\beta$, $\tau_{on}$, $p_{on}$, $\alpha$ and $x_{esc}$).
The excess kurtosis of all the PDFs shown in Fig. \ref{fig:ehist1} is negative, which implies that extreme values of the position and the velocity are less likely than in a Gaussian distribution.
For a larger amplitude of the active force, $\beta=10$, the PDFs of both the position and the velocity are very different from Gaussian and from the corresponding PDFs of trapped particles as shown in Fig. \ref{fig:ehist2}. Comparing Fig. \ref{fig:ehist2} with Fig. \ref{fig:nehist2} shows that the tri-modality found for the trapped particles in the position PDF does not appear for particles that can escape. On the other hand, for long auto-correlation times of the active force, tri-modality appears in the velocity PDF of particles that can escape, while the velocity PDF of trapped particles is uni-modal for the same parameters.
The fact that the PDF of the position shows no tri-modality is understood by the fact that particles escape if they reach $|x|=x_{esc}$. The tri-modality in the velocity is likely to be the result of the fact that under long periods of constant active force, the particles move with terminal velocity until they escape. The side peaks correspond to the terminal velocity in both directions (the active force is unbiased).

Figure \ref{fig:ehist3} shows that for weak confinement, $\alpha=0.0008$, a large amplitude of the active force, $\beta=100$, and a long auto-correlation time, $\tau_{on}=100t_d$, the position and velocity PDFs are strongly non-Gaussian, the excess kurtosis is large and positive (implying that extreme values are more likely than in a Gaussian distribution), and there are strong peaks for zero velocity and at the bottom of the potential well. The second moments of both the velocity and the position are smaller than those of trapped particles with the same parameters (see Fig. \ref{fig:nehist3}).

\begin{figure*}[b]
\includegraphics[width=\linewidth]{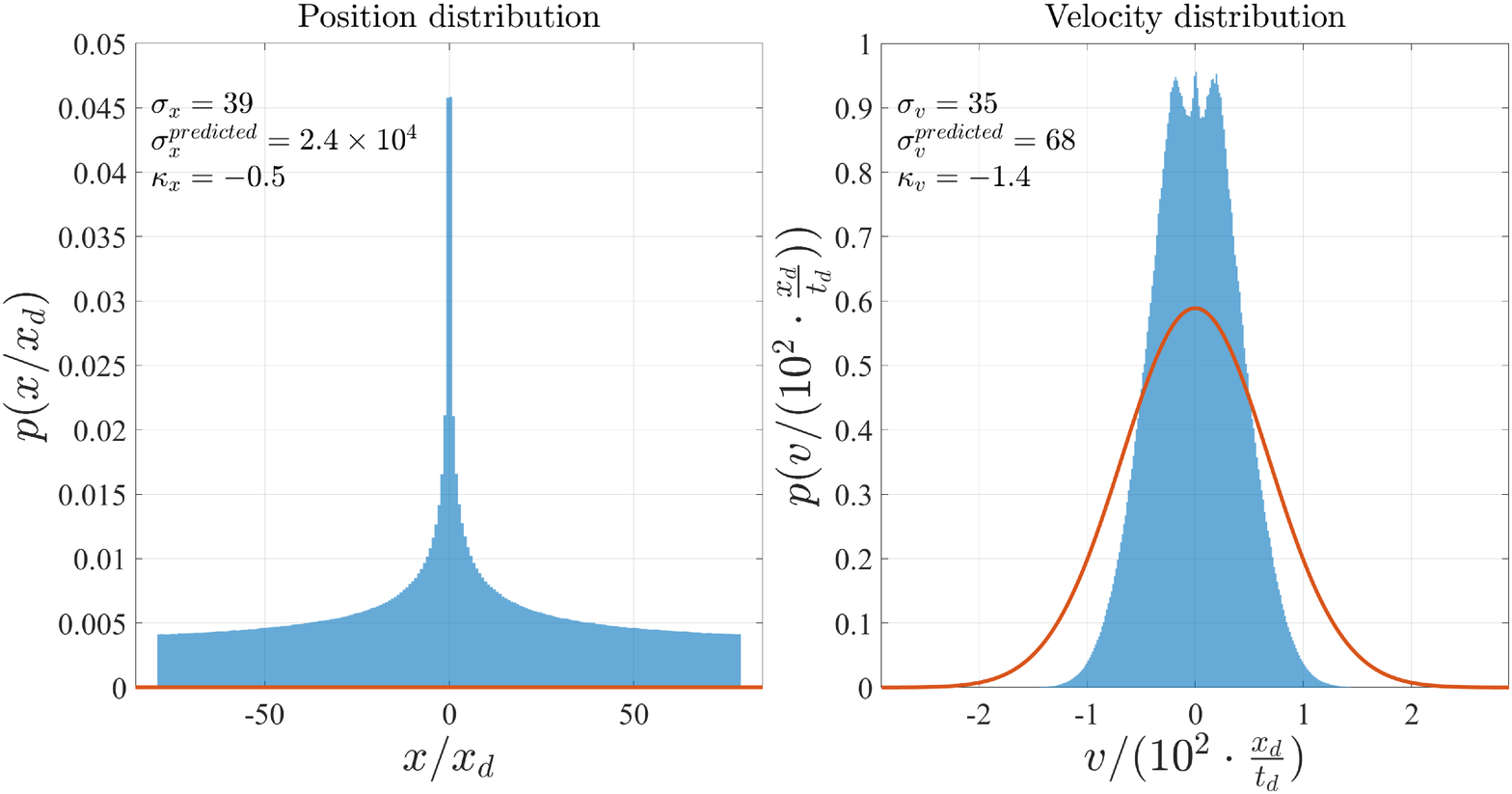}
\caption{\label{fig:ehistG3}
Position and velocity PDFs for a shallow trap, $\alpha=0.008$, with escape and Gaussian active noise ($\beta=100$ and $\tau_{on}=100t_d$). Due to the weak confinement, the PDFs are broader. The velocity PDF shows tri-modality which did not appear for the single source active noise. Moreover, both PDFs show negative excess kurtosis, while for the single source, they have large positive excess kurtosis. The escape point is $x_{esc}=x_d$.
}
\end{figure*}
\begin{figure*}[b]
\includegraphics[width=\linewidth]{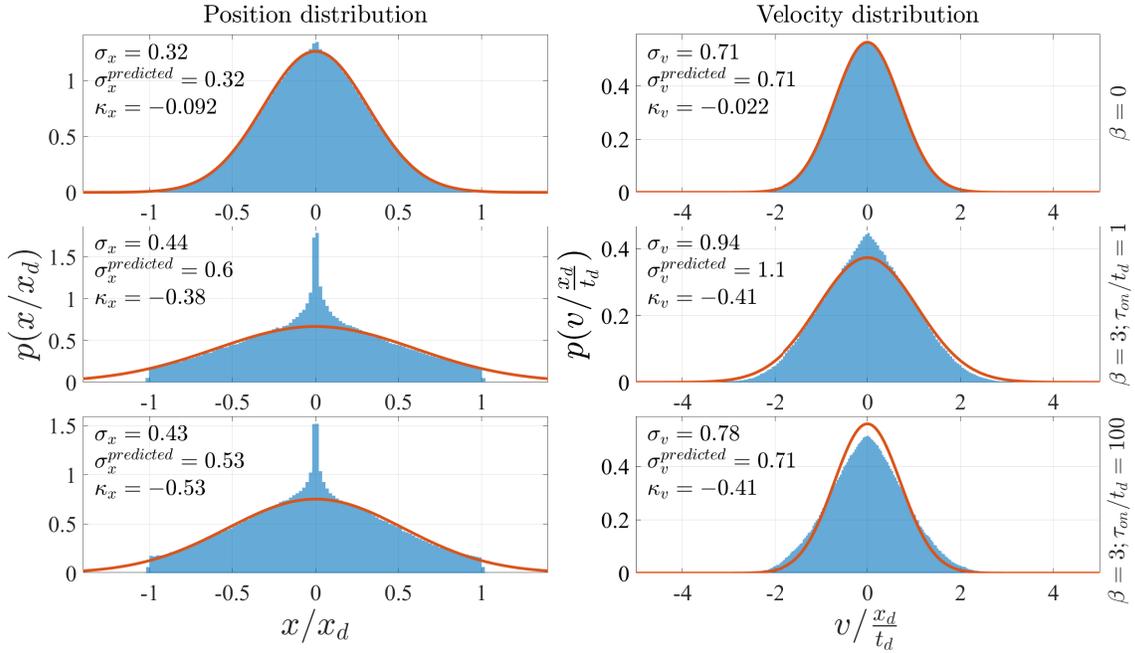}
\caption{\label{fig:ehistG1}
Position and velocity histograms for the case of escape and Gaussian active noise. The imposed vanishing probability density at the escape points sharply truncates the position PDF. The parameters are $\alpha=5$ and $x_{esc}=x_d$. The values of $\beta$ and $\tau_{on}$ are indicated in the figure.
}
\end{figure*}
\begin{figure*}[b]
\includegraphics[width=\linewidth]{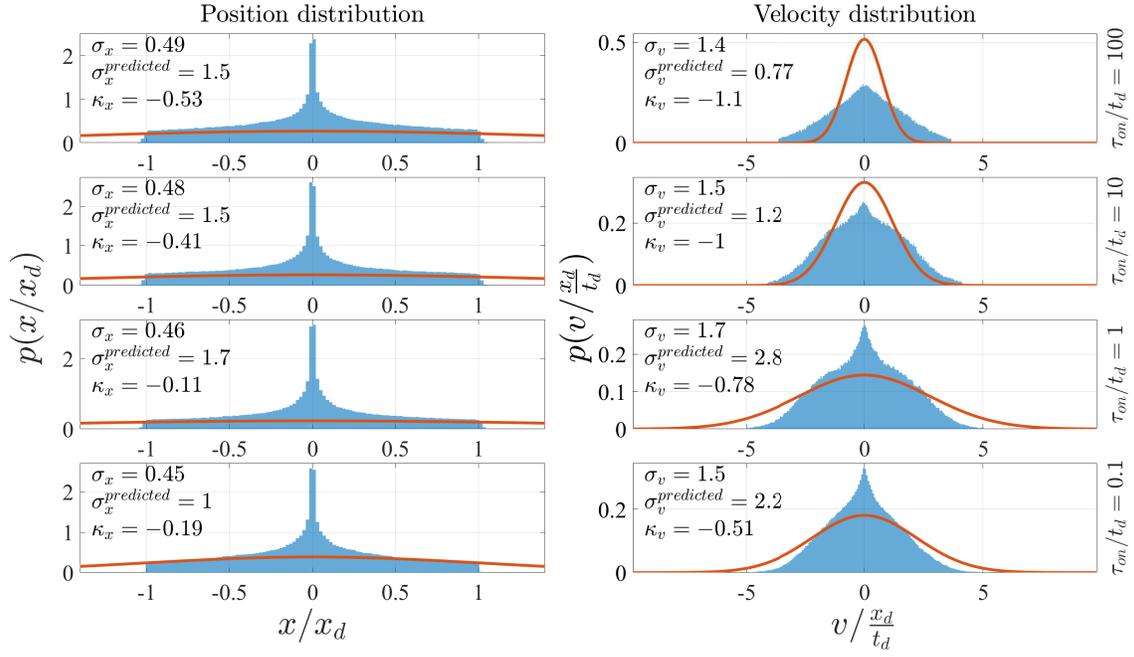}
\caption{\label{fig:ehistG2}
Similar to Fig. \ref{fig:ehistG1} but for $\beta=10$ and different values of $\tau_{on}$. The effect on the position PDF is similar to that seen in Fig. \ref{fig:ehist1}.
}
\end{figure*}

Figures \ref{fig:ehistG1}, \ref{fig:ehistG2} and \ref{fig:ehistG3} present the position and velocity PDFs for particles that can escape and are subjected to the Gaussian implementation of the active noise. For small amplitudes of the active force, Fig. \ref{fig:ehistG1}, the PDFs are similar to those found for the single source implementation of the force. For the larger amplitude, Fig. \ref{fig:ehistG2}, the position PDFs are similar to those obtained for the single source (Fig. \ref{fig:ehist2}), but the velocity PDFs show shoulders rather than the tri-modality obtained for the single source. This difference is due to the fact that for the Gaussian implementation, the force amplitude is not constant. For the large active force amplitude, we find apparent differences between the PDFs for the Gaussian noise, \ref{fig:ehistG3}, and those for the single source, Fig. \ref{fig:ehist3}. The position PDF shows negative excess kurtosis for the Gaussian noise, while it shows a large positive excess kurtosis for the single source. The velocity PDF shows the same difference, and in addition, for the Gaussian force, it shows tri-modality, which did not appear for the single source.

\clearpage
\renewcommand{\thefigure}{{D.\arabic{figure}}}
\section{Kramers' expression for the mean escape time}\label{App:tau0}
\setcounter{figure}{0}

The details of the pre-factor of the exponent, $\tau_{0}$, in Kramers' expression for the mean escape time, eq. \eqref{eq:Kramers}, depend on the shape of the potential and, in particular, on the way the potential is truncated (i.e., whether there is a cusp or just a smooth barrier).
The escape that we consider here corresponds to escape over a cusp for which the thermal Kramers' escape time, in the limit of moderate friction, is given by \cite{Kramer1,Kramer2,Melnikov1991},
\begin{equation}\label{eq:tau0mod}
    \tau_{esc,mf}^{thermal}=\left(2\pi/ \omega_0\right)e^{\Delta E(x_{esc})/(k_BT)},
\end{equation}
where $\omega_0=\sqrt{(1/m)|U''(x_0)|}$. In the limit of strong friction, Kramers' escape time is given by \cite{Kramer1,Kramer2,Melnikov1991},
\begin{equation}
    \tau_{esc,sf}^{thermal}=\frac{2\pi}{t_d\omega_0^2} \left(\frac{k_BT}{\pi\Delta E(x_{esc})}\right)^{1/2}e^{\Delta E(x_{esc})/(k_BT)}.
\end{equation}
The interpolation between moderate and strong friction was studied in \cite{Pollak1990}. In the limit of very weak friction, Kramers' mean escape time is independent of the barrier shape and is given by \cite{Kramer1,Kramer2,Melnikov1991}
\begin{equation}
\tau_{esc,wf}^{thermal}\approx t_d\frac{k_BT}{\Delta E}exp(\Delta E/(k_B T)).
\end{equation}

The results presented in Fig. \ref{fig:metvsb} illustrate the dependence of the mean escape time on the barrier height. The barrier was changed by either changing the escape point, $x_{esc}$, and fixing the stiffness of the confining potential, $\alpha$, or by changing the value of $\alpha$ and fixing the escape point. The curves for the thermal case show that even in the absence of active noise, when the escape point is fixed and the potential stiffness varies, the mean escape time deviates from exponential.

\begin{figure}[b]
\includegraphics[width=\linewidth,trim=4cm 0cm 4cm 2cm,clip]{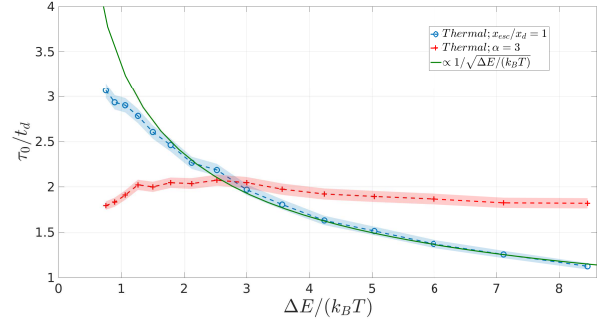}
\caption{\label{fig:tau0}
The pre-factor in Kramers' expression for the mean escape time vs. the energy barrier for the thermal (no active force) case of Fig. \ref{fig:metvsb}. The pre-factor shows the expected dependence on $\alpha$ for moderate friction and the lack of dependence on the escape point. The symbols denote the values of the energy barrier for which the mean escape time was simulated, and the shaded area represents the uncertainty.}
\end{figure}

In Fig. \ref{fig:tau0}, we present the numerical pre-factor of the exponent, $\tau_0$, that was obtained from the simulations. The energy barrier may be written as $\Delta E/(k_B T)=\alpha (x_{esc}/x_d)^2$. For the case of fixed $\alpha$, we find that $\tau_0$ only slightly varies and is, in fact, constant in the regime of high energy barriers, where Kramers' expression is expected to be valid. For the case of fixed escape point and varying stiffness, we found that $\tau_0\sim 1/\sqrt{\alpha}$, as expected in the regime of moderate friction. Since $x_{esc} =x_d= const$ (blue line and symbols), the energy barrier height is proportional to $\alpha$; the fit (green line) shows the predicted $1/\sqrt{\alpha}$ behavior (eq. \eqref{eq:tau0mod}).
%
% %
\clearpage

\renewcommand{\thefigure}{{E.\arabic{figure}}}
\begin{widetext}
\section{Effects of the active force correlation time on the mean escape time}\label{app:D}
\setcounter{figure}{0}
\end{widetext}

\begin{figure*}[b]
\includegraphics[width=\linewidth]{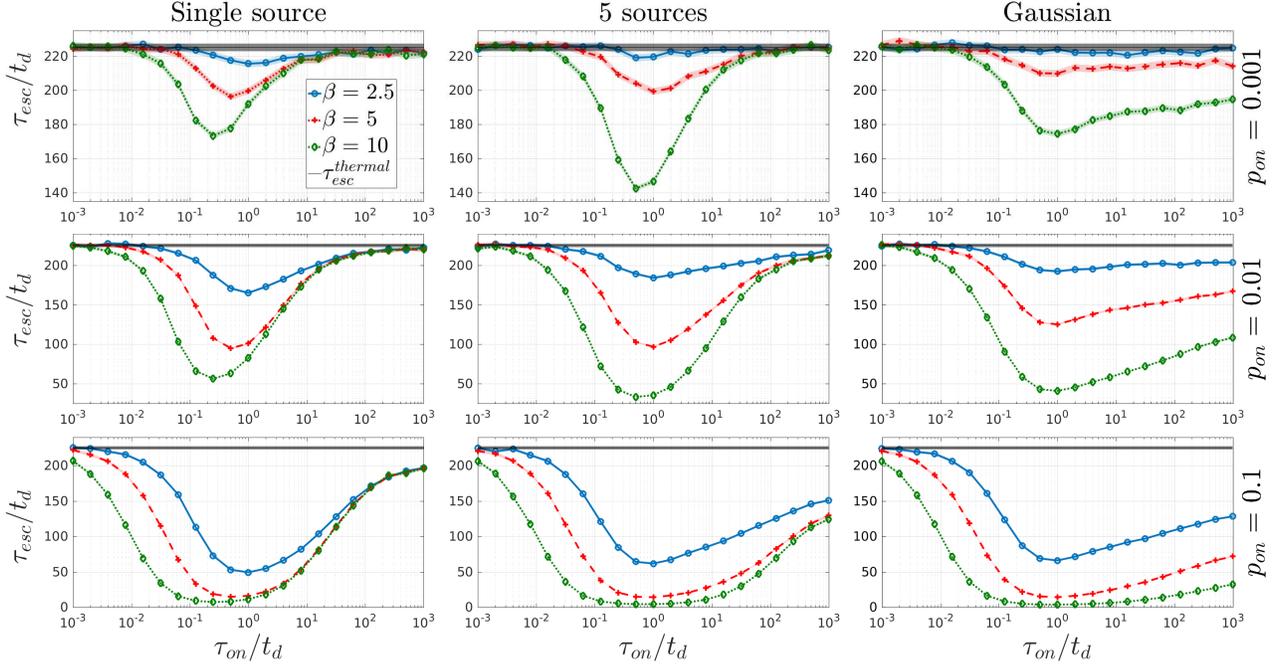}
\caption{\label{fig:tesc_cdc}
Mean escape time vs. the correlation time of the active force. The three rows correspond to the three indicated values of $p_{on}=\tau_{on}/\tau_{tot}$ (0.001, 0.01 and 0.1). The different columns correspond to three different implementations of the active force with a single source, five sources, and the Gaussian force, as indicated. Each panel includes results for three different amplitudes of the active force, $\beta$, (2.5, 5, and 10) as denoted. For all the simulations, $x_{esc}=x_d$ and $\alpha=5$. The black lines show the thermal escape time. The shaded areas represent the 95\% confidence interval. The confidence interval and the mean were derived from $4\times 10^4$ simulated trajectories.}
\end{figure*}
\begin{figure}[ht]
\includegraphics[width=\linewidth,trim=4.5cm 0cm 6cm 0cm,clip]{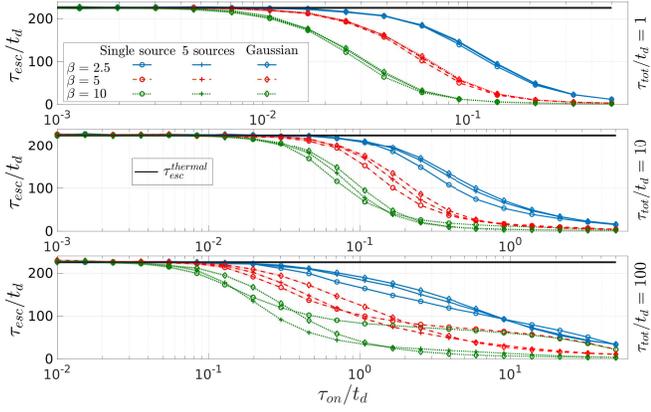}
\caption{\label{fig:tesc_cttot}
Mean escape time vs. the correlation time of the active force. Each panel corresponds to the indicated value of $\tau_{tot}$ (1, 10 and 100 in units of $t_d$). The different curves in each panel correspond to the nine combinations of three different force implementations (single source, five sources, and the Gaussian) and three amplitudes of the active force ($\beta=$2.5, 5 and 10).The shaded areas (too narrow to see in most cases) represent the 95\% confidence interval. The confidence interval and the mean were derived from $4\times 10^4$ simulated trajectories. For all the simulations, $x_{esc}=x_d$ and $\alpha=5$.}
\end{figure}
\begin{figure}[ht]
\includegraphics[width=\linewidth,trim=4cm 1cm 4cm 1cm,clip]{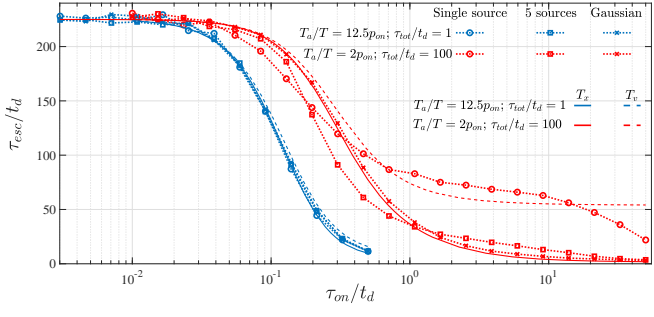}
\caption{\label{fig:tesc_comp}
Comparison of simulated mean escape time with theoretical approximation. Two distinct amplitudes of the active force were used, and for each of the three implementations of the active force. The effective temperature prediction's performance, for different regimes of the correlation time, can be seen.}
\end{figure}
\end{document}